\documentclass{aa}
\usepackage[varg]{txfonts}
\usepackage{amssymb}
\usepackage{amsfonts}
\usepackage{tabularx}
\usepackage{graphicx}
\usepackage{ulem}
\usepackage{array}
\usepackage{color}
\usepackage{booktabs}
\usepackage[fleqn]{amsmath}
\usepackage{subfigure}
\usepackage{hyperref}

\makeatletter
\newcommand*{\rom}[1]{\expandafter\@slowromancap\romannumeral #1@}
\makeatother
\numberwithin{equation}{section}
\def\d{{\rm d}}

\newcommand{\vv}{{\mathbf v}}
\newcommand{\vr}{{\mathbf r}}
\newcommand{\vR}{{\mathbf R}}
\newcommand{\vx}{{\mathbf x}}

\newcommand{\vA}{{\mathbf A}}

\renewcommand{\[}{\left[}

\renewcommand{\Re}{\textrm{Re}\,}

\begin{document}

\title{Effects of galaxy--satellite interactions on bar formation}

   \author{R. Moetazedian\inst{1}\thanks{Fellow of the International Max Planck Research School for Astronomy and Cosmic Physics at the University of Heidelberg (IMPRS-HD).}
          \and 
          E. V. Polyachenko\inst{2}
          \and
          P. Berczik\inst{1,3,4}
          \and
          A. Just\inst{1}\fnmsep\thanks{\email{just@ari.uni-heidelberg.de}}
          }

   \institute{Zentrum für Astronomie der Universit\"at Heidelberg, Astronomisches Rechen-Institut, M\"onchhofstr. 12-14, 69120 Heidelberg, Germany         
        \and
             Institute of Astronomy, Russian Academy of Sciences, 48 Pyatnitskya St., Moscow 119017, Russia
        \and
             Main Astronomical Observatory, National Academy of Sciences of Ukraine, MAO/NASU, 27 Akad. Zabolotnoho St. 03680 Kyiv, Ukraine
        \and   
             National Astronomical Observatories and Key Laboratory of Computational Astrophysics, Chinese Academy of Sciences, NAOC/CAS, 20A Datun Rd., Chaoyang District Beijing 100012, China
             }

   \date{Accepted on March 10, 2017}

 
  \abstract 
  {}
   {We aim to show how encounters with low-mass satellite galaxies may alter the bar formation in a Milky Way-like disc galaxy.}
   {We use high-resolution N-body simulations of a disc galaxy prone to mild bar instability. For realistic initial conditions of satellites, we take advantage of cosmological simulations of Milky Way-like dark matter haloes.}
   {The satellites may have a significant impact on the time of bar formation. Some runs with satellites demonstrate a delay, while others show an advancement in bar formation compared to the isolated run, with such time differences reaching $\sim 1$~Gyr. Meanwhile, the final bar configuration, including its very appearance and the bar characteristics such as the pattern speed and the exponential growth rate of its amplitude are independent of the number of encounters and their orbits. The contribution of satellites with masses below $10^9 M_{\odot}$ is insignificant, unless their pericentre distances are small. We suggest that the encounters act indirectly via inducing perturbations across the disc that evolve to delayed waves in the central part and interfere with an emerging seed bar. The predicted effect for the present-day host galaxy is expected to be even more significant at redshifts $z \gtrsim 0.5$.}
   {}
   
   \keywords{Galaxy: disk -- Galaxies: kinematics and dynamics -- Methods: numerical}
   \titlerunning{Bar-formation in the presence of satellites}            
   \maketitle

\section{Introduction} 
Structure formation in the Universe governed by the $\Lambda$ cold dark matter ($\Lambda$CDM) paradigm pursues a hierarchical model. Smaller DM (sub)haloes merge in order to build up larger objects (e.g. \citealt{white-rees,moore}). Such mergers occur on different scales, ranging from minor (1:3 - 1:50 of total mass) to major (mass ratios $\gtrsim$ 1:3 of total galaxy mass) mergers. These events are believed to impact the dynamics of galaxies, and even change their morphology (e.g. \citealt*{baugh,borne,hernandez,lotz10,kannan}). Current hydrodynamical cosmological simulations, which account for different small- and large-scale physical processes such as radiative cooling, feedbacks, star formation, magnetic fields and so on, are able to reproduce a spectrum of realistic galaxies with different morphologies, including ellipticals, barred, and unbarred spirals \citep{vogelsberger}.         

Observations of galaxies in the local and the high redshift Universe, have shown a high fraction of barred spirals (e.g. \citealt{bergh,abraham}). The properties of the host disc galaxy is affected by the presence of a bar, which possesses a strong non-axisymmetric gravitational potential. For instance, the redistribution of stars/gas within the bar's corotation radius, together with the evolution of the bulge, are strongly correlated with the bar (e.g. \citealt{hohl,kormendy82,athanassoula05}). Early N-body simulations were successful in generating spiral galaxies hosting bars (e.g. \citealt*{miller,ostriker,athanassoula86}).         

According to observations, a higher fraction of barred spirals are found in denser environments, such as galaxy groups and clusters, than in isolation, suggesting the importance of tidal interactions for bar formation \citep*{elmegreen}. \citet{noguchi87} pioneered numerical studies of bar formation in the presence of tidal interactions for initially stable galaxy models. In his runs, parabolic prograde planar very massive encounters with $M_\textrm{sat}/M_\textrm{galaxy} = 1 $ or 3 were able to only produce transient bars. Also, \citet{walker} showed that a satellite with $M_\textrm{sat}/M_\textrm{disc}=0.1$ on a circular prograde orbit with a pericentre distance of 6$R_\textrm{d}$ and inclined by 30$^\circ$ with respect to the disc plane is capable of destroying the bar in the initially bar-unstable models (their galaxies transformed into Sa Hubble morphological type). In \citet{moetazedian} we investigated the disc heating due to satellites infall in discs that were initially bar stable. In seven runs with different distributions of satellites extracted from zoom-in cosmological simulations of Milky Way-like hosts, no bar was induced, despite the mass ratios $0.003 < M_\textrm{sat}/M_\textrm{disc} < 4$. All these numerical experiments disfavour the tidally induced bar formation scenario in originally stable discs.

\citet{DBS08} discussed that massive satellite galaxies crossing the inner region of stable host galaxies may excite a bar or a spiral structure in the disc through the process of Toomre's swing amplification \citep[][hereafter, T81]{T81}, where the disc's inner part becomes vulnerable to such amplifications. 

\citet{kazan08} analysed the impact of massive encounters on the disc galaxy evolution in the cosmological context. For $z=1$ epoch, the disc mass is expected to be lower, and the satellites distribution contains more massive satellites with smaller pericentre distances compared to the present day epoch $z=0$. The adopted disc mass and the exponential scale length are $M_{\textrm d} = 3.53 \times 10^{10} M_\odot$ and $R_{\textrm d} = 2.82$~kpc, respectively. Also, the satellites were on eccentric orbits and possessed a mass range $0.21-0.57 M_{\textrm d}$ with pericentre distances of $0.5 -6.2 R_{\textrm d}$. Unfortunately, the authors do not provide information concerning the stability of the isolated run. The models with satellites show growth of strong central bars, and the disc axisymmetry is not restored at late times. We note that the Toomre stability parameter $Q=2.2$ adopted for the isolated model is insufficient to avoid local non-axisymmetric \citep{PPS97} and global instabilities such as bar instability as proved by our calculation of the isolated model with exactly the same Toomre parameter. 

For the host galaxy in the present analysis we employ a stellar dynamical model with a cuspy bulge generated using the GalactICS code~\citep{WPD08}, which was already studied in detail in our previous papers~\citep{PBJ16a, PBJ16b}. It turns out that the process of bar formation is affected by randomly incoming waves towards the disc centre, delaying it for some time, and then facilitating the bar growth \citep{P16}.

The analysis with different disc mass favours a young bar hypothesis, according to which the bar instability is saturated only recently~\citep{PBJ16b}. Thus, the present paper is aimed at investigating the importance of satellite galaxy encounters on the time of bar formation in a Milky Way-like galaxy using high-resolution N-Body simulations, with disc parameters and satellite distribution closer to the present day epoch, in contrast to \citet{kazan08}. In order to have realistic initial conditions (ICs), we use the distribution of satellites with masses exceeding $10^8 M_\odot$ extracted from cosmological simulations likely to host Milky Way-like galaxies \citep{aquarius,moetazedian}. 

In section~\ref{sec:nbody} we discuss the cosmological ICs employed together with the characteristics of our desired host galaxy and the satellite galaxies. Section~\ref{sec:inst-host} gives a short overview of bar instability in the host galaxy. The spectral analysis employed here follows that described by \citet{EP05}. The details of our simulations and the results obtained from our bar formation analysis are discussed in section~\ref{sec:results}. The relevance of the azimuthal phase between a weak bar and incoming satellites towards the bar formation epoch is investigated in section~\ref{sec:phase}. Also, the results from the run with an early satellite encounter are presented in section~\ref{sec:early}. Section~\ref{sec:observations} contains comparisons with observations. We finalise the paper with a summary and discussion of the results in section~\ref{sec:summary}. 

\section{N-body models}
\label{sec:nbody}
This section includes a description of the models recruited for setting up our desired Milky Way model, which consists of a disc, a bulge, and a DM halo together with the DM-only satellite galaxies.

\subsection{Cosmological initial conditions}
In order to account for a realistic image of satellite galaxies' infall onto the Milky Way's disc, the distribution of satellites was withdrawn from the Aquarius cosmological suite \citep{aquarius}. This consists of a set of six realisations of DM haloes likely to host Milky Way-like galaxies and which have not experienced a recent major merger. For the purpose of this study the level 2 of the Aquarius-D simulation, hereafter Aq-D2, was employed. We have shown in~\citet{moetazedian} that Aq-D2 is a fair representative of a typical Milky Way-like system with contribution towards the vertical heating of the Galactic disc regarded as average compared to the rest of Aquarius simulations.

In our study we use the $z=0$ snapshot which corresponds to the present day distribution of DM substructures. The snapshot contains information such as position, velocity, maximum circular velocity $V_\textrm{max}$, the radial distance of this velocity $r_\textrm{vmax}$ and the tidal mass $M_\textrm{tid}$ for every substructure (satellite) within the simulation box. The parent host halo has an enclosed mass, $M_\textrm{200}=1.774\times10^{12} M_{\odot}$, which corresponds to the mass within a sphere with radius $r_\textrm{200}=242.8$~kpc; this represents the radius at which the mean density of the DM halo is 200 times the critical density of the Universe ($\rho_\textrm{crit}$). Also the radius, where the halo encloses the mass with mean density 50 times $\rho_\textrm{crit}$ is denoted as $r_\textrm{50}$ and has a value of 425.7~kpc.

As mentioned earlier we are interested in satellites with a chance of passing close to the  disc. The detailed procedure of identifying these candidates, is described in a previous paper~\citep{moetazedian}. The statistics of the satellite distribution is shown in Tab.~\ref{tab:2}; $N_\textrm{sub}$ is the total number of satellites with $M_\textrm{tid}\geq10^{6} M_{\odot}$ and $f_\textrm{sub}$ ($< r_\textrm{50}$) is the fraction of these objects within $r_\textrm{50}$. The percentage fraction of crossed satellites (i.e. objects that come closer than 25~kpc to the host halo's centre during their 2~Gyr orbit) is represented by $f_\textrm{cross}$ ($< r_\textrm{50}$). The last four rows show the number of crossed satellites with masses $\geqslant$ 1, 3, 5 and $10\times10^{8} M_{\odot}$. We argue in~\citet{moetazedian} that only satellites with $M_\textrm{tid}\geqslant10^{8} M_{\odot}$ potentially contribute towards the heating of the Galactic disc. The Aq-D2 simulation has 23 satellites with $M_\textrm{tid}\geqslant10^{8} M_{\odot}$ out of which four have masses larger than $10^{9} M_{\odot}$. 
 \begin{table}
 \caption{Number statistics of Aq-D2 satellites.}
 \label{tab:2}
 \begin{tabularx}{\linewidth}{XX}
  \hline
 Quantity & Value \\ \hline
 $N_\textrm{sub}$ ($M_\textrm{tid}\ge10^{6} M_{\odot}$) & 72,380\\
 $f_\textrm{sub}$ ($<$ $r_\textrm{50}$) & 24.01 \% \\
 $f_\textrm{cross}$ ($<$ $r_\textrm{50}$) & 8.85 \% \\
 $N_\textrm{cross}$ ($>$ 10$^{8}$ $M_{\odot}$) & 23 \\
 $N_\textrm{cross}$ ($>$ 3 $\times$ 10$^{8}$ $M_{\odot}$) & 8 \\
 $N_\textrm{cross}$ ($>$ 5 $\times$ 10$^{8}$ $M_{\odot}$) & 5 \\
 $N_\textrm{cross}$ ($>$ 10$^{9}$ $M_{\odot}$) & 4 \\ \hline
 \end{tabularx}
 \tablefoot{$f_\textrm{sub}$ ($< r_\textrm{50}$) is the percentage of the satellites originally within $r_\textrm{50}$ of the total number of satellites $N_\textrm{sub}$ with $M_\textrm{tid}\geq10^{6} M_{\odot}$, while $f_\textrm{cross}$ shows the percentage of satellites inside $r_\textrm{50}$ which cross the disc in 2~Gyr. $N_\textrm{cross}$ give the numbers of crossed satellites above the noted threshold.}
 \end{table} 

The host halo, in which these satellites would be inserted, has a slightly different enclosed mass ($M_{200}$). Therefore, we needed to rescale the phase-space properties of the satellites, together with their masses, following a recipe mentioned by~\citet{kannan}, in order to have a physically sensible analysis. The scaling factor is defined as \textit{f}=$M_\textrm{200,Aq-D2}/M_\textrm{200}$ (=1.37 in our case),

\begin{equation}
  M=M_\textrm{orig}/f,
  \label{eq:scl1}
\end{equation} 
\begin{equation}
  v_\textrm{\textit{x,y,z}}=\frac{v_{\textrm{orig},x,y,z}}{f^{1/3}} \qquad \textrm{and}
  \label{eq:scl2}
\end{equation} 
\begin{equation}
  x=\frac{x_\textrm{orig}}{f^{1/3}}
\qquad
  y=\frac{y_\textrm{orig}}{f^{1/3}}
\qquad
  z=\frac{z_\textrm{orig}}{f^{1/3}}.
  \label{eq:scl3}
\end{equation} 
The subscript ``orig'' represents the original non-scaled values.
Early CDM N-body simulations have shown that the density profile of DM (sub)haloes could be well fitted via the known Navarro-Frank-White (NFW) profile~\citep{nfw} 
\begin{equation}
  \rho_\textrm{NFW}(r)=\frac{\rho_\textrm{s}}{(r/r_\textrm{s})(1+r/r_\textrm{s})^2}
  \label{eq:nfw}
\end{equation}
and
\begin{equation}
 \delta_c=\frac{\rho_\textrm{s}}{\rho_\textrm{crit}} = \frac{200}{3}\frac{c^{3}}{\ln(1+c)-c/(1+c)}\ .
  \label{eq:deltac}
\end{equation}
Here $\rho_\textrm{s}$ and $r_\textrm{s}$ are the scale density and the scale radius of the halo, respectively. The density contrast $\delta_c$ with respect to $\rho_\textrm{crit}$ is calculated  with the concentration $c$ of the halo being the ratio of $r_\textrm{200}/r_\textrm{s}$.

As discussed in section~\ref{sec:sat}, all the DM satellites also follow a NFW profile, which is tidally truncated at $r_\textrm{tid}$ according to the satellite mass $M_\textrm{tid}$.

\subsection{The host galaxy}
Our three-component model is adopted from~\citet{WPD08, KD95}, and consists of a stellar disc, a bulge, and a DM halo. The complete list of characteristics for the disc, the bulge and the halo components are listed in Tab.\,\ref{tab:4}. For a recent discussion of global parameters of the Milky Way see \citet{BG16}.

The disc is exponential, with radial scale length $R_\textrm{d}=2.9$~kpc and truncation radius 15~kpc. The radial velocity dispersion $\sigma_R$ is approximately exponential with central value $\sigma_{R0}=140$~km\,s$^{-1}$ and radial scale length $R_{\sigma}=2R_\textrm{d}$. The solar neighbourhood location is at $R=8$~kpc (e.g. \citealt{gillessen,reid}). 

\begin{table}
\caption{The characteristics of disk+bulge+halo for our host galaxy model.} 
\label{tab:4}
\begin{tabularx}{\linewidth}{XX}
 \hline
Quantity & Value \\ \hline
$\Sigma_\textrm{sol}$ & 50 $M_{\odot}$pc$^{-2}$\\
$R_\textrm{d}$ & 2.9 kpc \\
z$_\textrm{d}$ & 300 pc \\
$M_\textrm{d}$ & 4.2 $\times$ 10$^{10}$ $M_{\odot}$ \\
$\sigma_{R}$ & 35 km\,s$^{-1}$ \\
$\sigma_{R0}$ & 140 km\,s$^{-1}$ \\
$N_\textrm{d}$ & 6 million \\ [2mm]
$R_\textrm{e}$ & 0.64 kpc \\
$\sigma_\textrm{b}$ & 272 km\,s$^{-1}$ \\
$M_\textrm{b}$ & 1.02 $\times$ 10$^{10}$ $M_{\odot}$ \\
$N_\textrm{b}$ & 1.5 million \\[2mm]
$M_{200}$ & 1.29 $\times$ 10$^{12}$ $M_{\odot}$ \\
$r_\textrm{s}$ & 17.25 kpc \\
$r_\textrm{200}$ & 229.3 kpc \\
$N_\textrm{h}$ & 9.25 million \\ \hline
\end{tabularx}
\tablefoot{$N$ represents the particle number for the component and $M$ the mass, while the subscripts d, b, and h correspond to disc, bulge, and halo. The solar neighbourhood surface density $\Sigma_\textrm{sol}$ together with the thin disc scale height and scale length, z$_\textrm{d}$ and $R_\textrm{d}$ characterize the disc density distribution. $\sigma_{R}$ and $\sigma_{R0}$ correspond to the solar and central radial velocity dispersions of the disc. In the case of the bulge, $R_\textrm{b}$ and $\sigma_\textrm{b}$ are the effective radius and the characteristic velocity scale.}
\end{table} 

The bulge component takes a density profile of the following form 
\begin{equation}
\rho_\textrm{b}(r) = \rho_\textrm{b} \left( \frac r{R_e} \right)^{-p} \textrm{e}^{-b(r/R_e)^{1/n} }\ ,
\label{eq:bulge_dens}
\end{equation}
where $r$ is the spherical radius, corresponding to a S\'{e}rsic surface brightness profile with effective radius $R_e$.
The scale density $\rho_\textrm{b}$, can be replaced by the bulge velocity scale 
\begin{equation}
\sigma_\textrm{b} \equiv \left\{ 4\pi G n b^{n(p-3)} \Gamma[n(3-p)] R^2_e \rho_\textrm{b} \right\} ^{1/2}\ .
\label{eq:bulge_veld}
\end{equation}
Here, $\sigma^2_\textrm{b}$ corresponds to the depth of the gravitational potential associated with the bulge. In addition, we require $n=1.11788$ leading to  $p\simeq0.5$ parameter, in order to define the bulge density profile.

We have employed a truncated NFW profile for the halo, with the truncation radius at $r_\textrm{200}$. The bulge component dominates the inner rotation curve at $R \gtrsim 0.01$~kpc, despite the halo having a more cuspy profile. 

\begin{figure} 
\centering
  \centerline{\includegraphics[width = \linewidth]{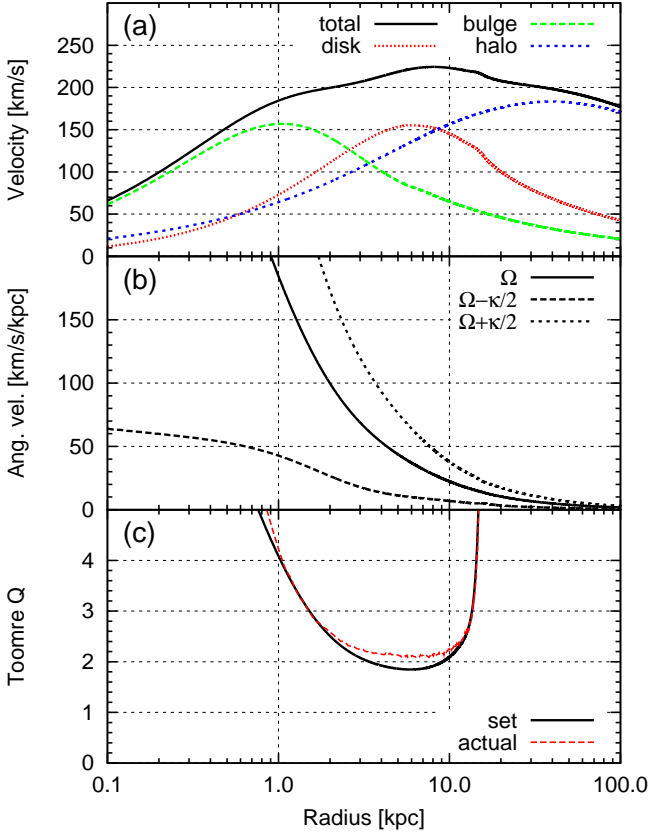}}
  \caption{Initial profiles for the basic model: (a) the total circular velocity and its components due to disc, bulge, and halo; (b) angular velocity $\Omega(R)$ and curves $\Omega(R) \pm \kappa(R)/2$; (c) Toomre $Q$ initially set by GalactICS  and the one actually obtained in the simulations.}
  \label{fig:1}
\end{figure}

The total circular velocity profile (solid curve) and the contribution from each component (dashed/dotted) are shown in the top panel (a) of Fig.~\ref{fig:1}. The bulge dominates at radii $R\lesssim 1.5$~kpc, while the halo takes over at $R > 10$~kpc. At $R\approx 6$~kpc, where the contribution of the disc component reaches the maximum, the force from the halo is approximately two thirds of that from the disc in the Galactic plane. 

Panel (b) presents the angular velocity profile $\Omega(R)$ in the equatorial plane. In addition, positions of the inner (ILR) and outer Lindblad resonances (OLR) are shown as $\Omega \pm \kappa/2$. For a given pattern speed $\Omega_\textrm{p}$, the ILR is calculated using
\begin{equation}
\Omega_\textrm{p} = \Omega_\textrm{pr}(R) \equiv \Omega(R) - \frac12 \kappa(R) \ .
\label{eq:res}
\end{equation}
The quantity, $\Omega_\textrm{pr}(R)$ is responsible for determining a precession rate of nearly circular orbits. Hence, it can be referred to as `precession' curve, which diverges weakly as $R \to 0$ with $R^{-\alpha/2}$ and $\alpha \approx 0.5$. 

The lower panel (c) of Fig.\,\ref{fig:1} demonstrates the Toomre $Q$ profile
\begin{equation}
Q = \frac{\kappa \sigma_R}{3.36 G\Sigma_\textrm{d}}\ ,
\label{eq:Qs}
\end{equation}
where $\Sigma_\textrm{d}$ represents the disc surface density. The solid line shows the profile which was initially set by the GalactICS code, while the dashed curve corresponds to the actual measured profile in our Galaxy run. The deviation for $2 <R<12$ kpc is explained by larger actual radial dispersion. The initial minimum $Q_\textrm{min}=1.8$ resides at $R=5.9$~kpc. This minimum rises to $Q_\textrm{min}\approx 2.1$ in case of the actual run.

\subsection{Satellites}
\label{sec:sat}
In order to insert the satellites into our N-body simulations, each satellite needs to be generated as a distribution of particles following their corresponding NFW profile. 
The parameters $\rho_\textrm{s}$ and $r_\textrm{s}$ in Eq.~\ref{eq:nfw} are determined by  $r_\textrm{vmax}$ using $r_\textrm{s}=0.4623\,\,r_\textrm{vmax}$  and $V_\textrm{max}$ using the enclosed mass inside $r_\textrm{vmax}$. Next we derive the  tidal radius $r_\textrm{tid}$ as cut-off radius from the cumulative mass profile reaching the satellite mass $M_\textrm{tid}$.

The distribution function introduced by~\citet{WPD08} and implemented in~\citet{lora} was used for generating cuspy NFW profiles. Such a profile has an infinite cumulative mass at $r \rightarrow \infty$; therefore, we use the satellites' $r_\textrm{tid}$ as the cut-off radius. In this work, all satellites are represented using 50,000 particles. 

\begin{figure}
\centering 
  \centerline{\includegraphics[width = \linewidth]{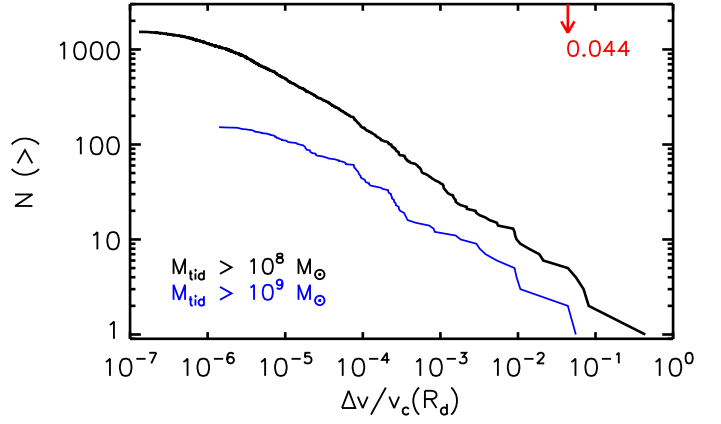}}
  \caption{The reverse cumulative histograms of the number of encounters with the tidal impact parameter (\ref{eq:dV}) greater than the given one, for all seven runs of \citet{moetazedian}. We consider the encounters with $M_{\textrm{tid}} > 10^8 M_\odot$ (black) and $M_{\textrm{tid}} > 10^9 M_\odot$ (blue) before applying the rescaling. The red arrow marks the position of the Aq-D2 encounter used in simulations with one satellite and it is the 5th encounter from the right hand-side of the upper cumulative line. The satellite with the highest impact ($\sim0.4$) has a mass $1.1\times10^8 M_\odot$ and comes as close as 0.25~kpc to the centre.}
  \label{fig:sat_stat}
\end{figure}

The satellites can be ranged according to their ability to excite density waves in the disc of the host galaxy. In order to quantify it, we shall introduce a `tidal impact' parameter, which is a normalised tidal velocity perturbation. For a crude estimate, we use the formulae for tidal shocks \citep[][sect. 8.2.1]{BT08}, calculating the ratio of the perturbation to the circular velocity at some typical radius, for example, $v_{\textrm c}(R_{\textrm d}) $:
\begin{equation}
\left( \frac{\Delta v}{v_{\textrm c}}\right)_{R_{\textrm d}} \sim \frac{2 G M_{\textrm{tid}}}{b^2 V} \frac{R_{\textrm d}}{v_{\textrm c}} \sim \frac{G M_{\textrm{tid}} R_{\textrm d} }{b^2 v^2_{\textrm c}}  \ .
\label{eq:dV}
\end{equation}
Here $b$ is the pericentre distance and we assumed that velocities at the time of encounters $V$ are similar and approximated as $V \simeq 2 v_{\textrm c}$. Fig.\,\ref{fig:sat_stat} shows the reverse cumulative histogram of the number of encounters with the tidal impact parameter (\ref{eq:dV}) greater than a given one, for all encounters with $M_{\textrm{tid}} > 10^8 M_\odot$ among seven runs discussed in \citet{moetazedian}. The Aq-D2 run can be regarded as a natural choice, since it contains at least a couple of satellites with tidal impact parameters at the high end. The marked red arrow corresponds to the Aq-D2 satellite with the highest parameter (referred to hereafter as `primary'), coming as close as $\sim4$~kpc to the centre, and having the second highest mass in Aq-D2. The primary satellite is expected to have a larger tidal impact on the disc, by a factor of $\sim2$, compared to the most massive satellite. 

The host halo has a different enclosed mass and concentration compared to the Aq-D2 main halo. This means the orbits of Aq-D2 satellites would differ if inserted into this model. In order to gain fair and physically sensible results, we decided to scale $M_\textrm{tid}$, positions and velocities of Aq-D2 satellites using the factor $f=1.37,$ which corresponds to the ratio of Aq-D2 and the targeted host halo mass. Simulations with these initial conditions are called `fully-rescaled'. For comparison we also performed simulations where only the masses of the satellites are rescaled, but not their initial orbital positions and velocities, which we call `mass rescaled' (see Tab.\,\ref{tab:runs}).

\section{Bar instability in the host galaxy}
\label{sec:inst-host}
The isolated galaxy is prone to instability leading to the formation of a bar. Bar properties on low amplitudes, such as a pattern speed and an amplitude (exponential) growth rate, can be obtained as global solutions from matrix equations \citep[e.g.][]{K71, K77}, describing a disc with stellar orbits of different eccentricities. It opposes tightly wound local WKB solutions of Lin--Shu--Kalnajs obtained in the epicyclic approximation.

The applicability of the razor-thin disc model for the description of real discs with finite thickness and a density cusp in the centre has been studied in detail by \citet{PBJ16a}. The main idea is to consider an effective rotation curve, which takes into account the $z$-dependence of the radial force on the stars that elevate above the equatorial plane. The effective rotation curve lacks cuspy features and possesses a maximum, which allows for a bar in a wide range of pattern speeds.

Here we use a matrix method by \citet{EP05} that has a form of the linear matrix equation,
\begin{equation}
    \vA \vx = \omega \vx\ ,
    \label{eq:me}
\end{equation}
which allows us to find unstable modes effectively without prior information on the localisation of modes.

A serious flaw in all matrix methods is the inability to calculate unstable global modes in a live halo and bulge. Although the substitution of a live halo with its rigid counterpart gives nearly the same pattern speeds as in the fully live models, the growth rates are smaller by a factor of two or even more. The theoretical bar pattern speed estimate for our host galaxy is  $\Omega_\textrm{b} \approx 48.34$ km\,s$^{-1}$\,kpc$^{-1}$ and for the growth rate we obtained $\omega_\textrm{I} \approx 1.23$~Gyr$^{-1}$, so the expected value for the fully live models is 2...3~Gyr$^{-1}$.

\section{N-body simulations}
\label{sec:results}
The initial conditions of stars for single mass simulations have been generated by the `GalactICS' code provided by~\citet{WPD08}. In our previous N-body simulations~\citep{PBJ16a} we used three different codes ({\tt SUPERBOX--10}, {\tt bonsai2}, {\tt ber-gal0}) and demonstrated the practical independence of the results on the chosen code. 
The simulations in this work are restricted to the {\tt bonsai2} tree-code and {\tt ber-gal0}\footnote{\tt ftp://ftp.mao.kiev.ua/pub/users/berczik/ber-gal0/}~\citep{ZBGPJ2015} routines as an auxiliary tool to analyse the snapshot data. The modified version of the recently developed N-body Tree--GPU code implementation {\tt bonsai2}~\citep{BGPZ2012a, BGPZ2012b} includes expansion for force computation up to quadrupole order. The opening angle used had a value of $\theta=0.5$, accompanied by individual gravitational softening of 10~pc. {\tt bonsai2} employs the leap-frog integration scheme with a fixed time--step $\Delta t=0.2$~Myr.

The current set of simulations were carried out with the GPU version of the code using the ARI GPU cluster {\tt kepler} and also the GPU cluster {\tt MilkyWay} specially dedicated to the SFB 881 (``The Milky Way System''), located in the J\"ulich Supercomputing Centre in Germany.

Tab.\,\ref{tab:runs} contains the summary of our runs. The capital letter `B' denotes the numerical \textsc{bonsai2} Tree code. It is worthwhile to mention, the Milky Way halo components have multi-mass halo particles in order to achieve a better resolution in the bar region. For this, we modified the GalactICS code utilising the same strategy as described in \citet{DBS09}. In the region between 0.1 and 1~kpc, the number density ratio of our multi-mass and single mass runs varies from 10 to 100, thus the effective numerical resolution there is enhanced by this factor.

\begin{table}
\begin{center}
\caption{Summary of the runs}
\label{tab:runs}
\begin{tabular}{l l l l l l l l l}
\hline
 Model & $n_\textrm{S}$ & $\Delta_\textrm{S}$ & $t_1$ &  $t_2$ &  $t_3$ & $\Delta_B$  &  $\Omega_\textrm{p}$ & $\omega_\textrm{I}$ \\
\hline
B-0 & --               & -- &  0.75 & 1.4 & 2.2 & -- & 49 & 2.9 \\[2mm] 
B-1 & 1                & -- &  0.75 & 1.2  & 1.8 & -200 & 48 & 3.0 \\ 
B-1m & 1               & -- &  0.75 & 2 & 2.4 & 300 & 49 & 2.9 \\ 
B-4 & 4                & -- &  0.75 & 1.3 & 2.0 & -300 & 49 & 2.9 \\ 
B-4m & 4               & -- &  0.75 & 1.7 & 2.5 & 470 & 48 & 3.0 \\ 
B-23 & 23              & -- &  0.75 & 1.1 & 1.7 & -310 & 48 & 2.9 \\ [2mm]
B-1-110 & 1          & +110 &  0.75 & 1.6 & 2.5 & 600 & 48  & 2.9 \\ 
B-1-500 & 1       & --\,500 &  0.40 & 1.0 & 2.2 & -20 & 48 & 2.7\\ 
\hline
\end{tabular}
\end{center}
\vspace{-2mm}
\tablefoot{$n_\textrm{S}$ is the number of satellites; `m' denotes only mass-rescaled satellites while the remaining simulations are fully -- mass, positions and velocities -- rescaled. $t_1$, $t_2$, $t_3$ denote the times (in Gyr) of the jump, the onset and the end of the exponential growth of the bar, respectively. $\Delta_B$ is the delay of the bar growth in Myr with respect to the isolated case B-0. Pattern speed $\Omega_\textrm{p}$ in km\,s$^{-1}$ and exponential growth rate $\omega_\textrm{I}$ in Gyr$^{-1}$ are given in the last columns.
}
\end{table}

The subscript `m' corresponds to the simulations, which are only mass-rescaled using Eq.~\ref{eq:scl1}; while the remaining simulations are fully-rescaled. In the case of simulations marked using `B-1', we use the primary satellite from Aq-D2 simulation with rescaled mass $M_\textrm{tid}=2.45\times10^{9} M_{\odot}$. For runs marked as `B-4', the four most massive Aq-D2 crossed satellites with $M_\textrm{tid} > 10^{9} M_{\odot}$ were employed.

\begin{figure}
\centering 
  \centerline{\includegraphics[width = 85mm]{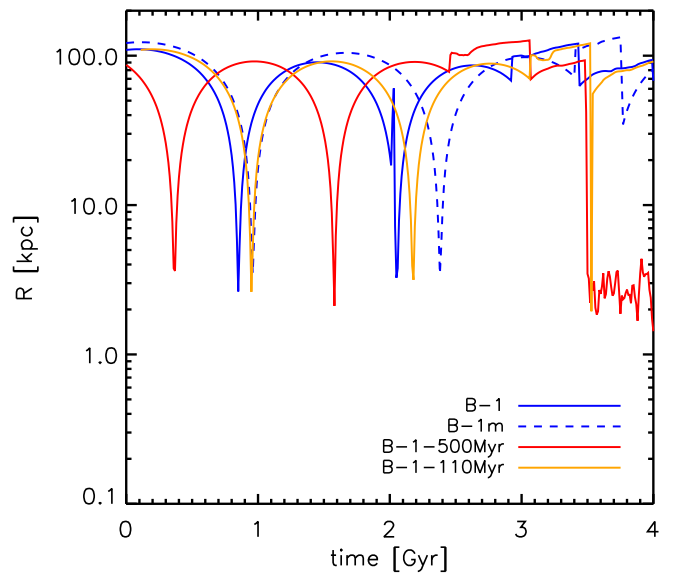}}
  \caption{The orbit of the primary satellite from our five runs as a function of time.}
  \label{fig:sat_orbit_bs}
\end{figure}

The orbital positions and velocities of the satellites are determined by their density centres. Fig.~\ref{fig:sat_orbit_bs} shows the galactocentric orbital distance for Aq-D2's primary satellite from our five runs. There exists a 110~Myr difference between the first pericentre passage of B-1 and B-1m runs. The pericentre distances have the values of 2.6 and 3.5~kpc, occurring at 0.85 and 0.96~Gyr, respectively. Also, the orbits from the two simulations with $+110$ and $-500$~Myr time shift are shown.

\subsection{Isolated simulation B-0 of the host galaxy}
\begin{figure}
\centering
\centerline{\includegraphics [width = \linewidth]{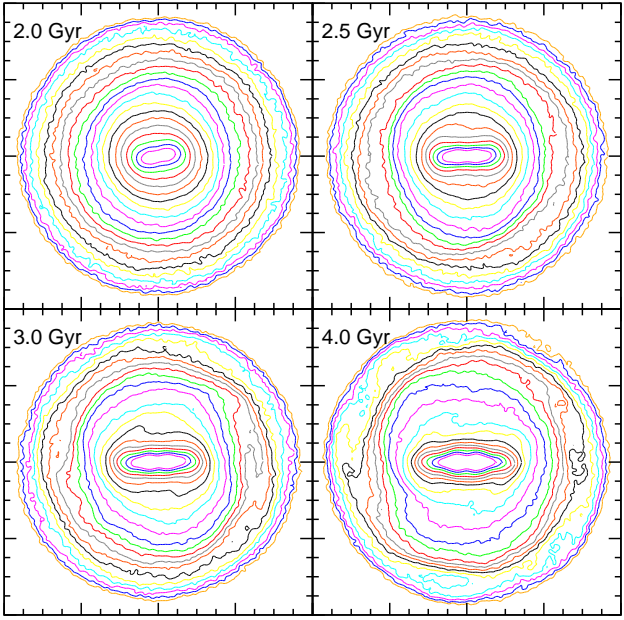}}
\caption{Bar patterns in the B-0 run oriented along the $x$-axis at different stages of bar evolution. The curves are isolines of the surface density evenly spaced in log scale (ten levels for every factor of 10). Each frame size is $32\times32$~kpc.} 
\label{fig:gS1}
\end{figure}

In the absence of the satellites, the host galaxy appears unstable as manifested in the formation of a bar. In Fig.\,\ref{fig:gS1} one can see snapshots at times 2, 2.5, 3, and 4~Gyr of the bar oriented along the $x$-axis. At earlier times, the asymmetry of the density distribution is weak, because of a delay in bar formation (e.g.~\citealt{PBJ16a,P16}). Moreover, there are no noticeable spirals throughout the simulation.  

The bar is a wave, and its growth can be investigated using relative amplitudes of perturbed quantities, such as the surface density $A_2/A_0$,
\begin{equation}
A_m(t) = \sum\limits_{j} \mu_j \textrm{e}^{-im\theta_j}
\label{eq:A2A0}
,\end{equation}
(here $\mu_j$ and $\theta_j$ are mass and polar angle of star $j$; $j$ spans particles within some fixed radius, for example, the radial scale length $R_\textrm{d}$), or velocity components $U_{k,m}$. The latter are defined as the perturbed velocity amplitudes,
\begin{equation}
V_{k,m}(t) = \frac{1}{A_0} \sum\limits_{j} \mu_j \tilde v_{k,j} \textrm{e}^{-im\theta_j}\ ,
\label{eq:Vkm}
\end{equation}
normalized to the average velocity dispersion $\sigma_k(t)$ calculated by
\begin{equation}
\sigma^2_k(t) = \frac{1}{A_0} \sum\limits_{j} \mu_j \tilde v^2_{k,j}\ .
\label{eq:sigma}
\end{equation}
The cylindrical components $R$, $\theta,$ and $z$ correspond to the quantity $k,$ and $\tilde v_{k,j}$ is the residual velocity component of the particle $j$. 

\begin{figure}
\centering
   \centerline{\includegraphics [width = \linewidth]{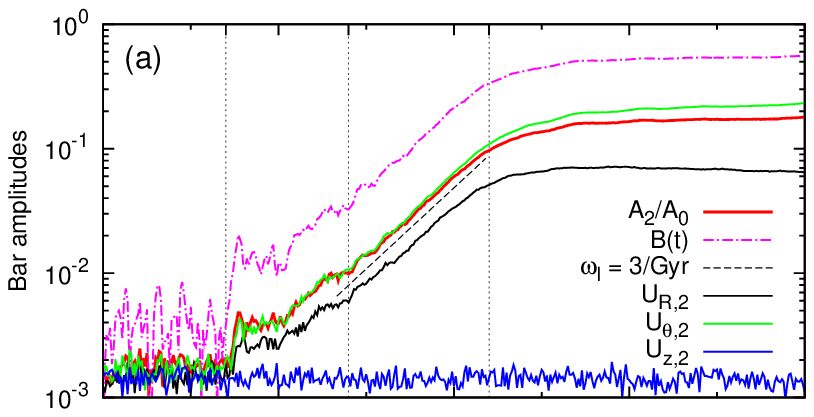}} 
\vspace{-7mm}
   \centerline{\includegraphics [width = \linewidth]{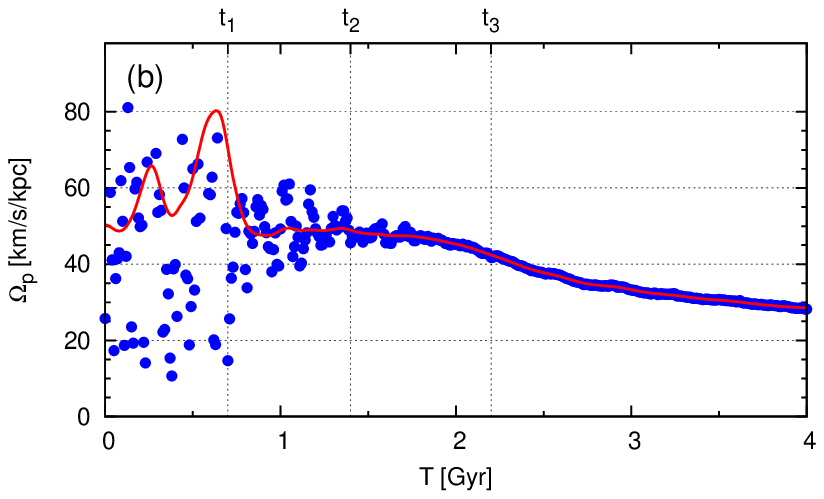}}
   \caption{Bar formation in the isolated model. (a) bar amplitudes $A_2/A_0$ and $U_{R\theta,2}$, and the bar strength $B(t)$ demonstrate lags, jumps, and exponential growths with a rate $\omega_\textrm{I} \approx 3$~Gyr$^{-1}$ (the slope is shown by the black dashed curve). (b) the pattern speed of the bar $\Omega_\textrm{p}$ determined as an average centred finite difference of the bar phase (dots), and the spline smoothed $\Omega_\textrm{p}$ (red curve).}
   \label{fig:bs_live}
\end{figure}

Panel (a) in Fig.~\ref{fig:bs_live} shows the relative bisymmetric amplitudes $A_2/A_0$, and $U_{k,2}$ for the B-0 run, representing the isolated galaxy. All the curves, except $U_{z,2}$, 
show a typical phase of low amplitude fluctuations, followed by a jump. The evolution then enters a phase characterised by a typical exponential growth, and finally reaching a plateau. In Tab.~\ref{tab:runs}, $t_1$ denotes the jump, $t_2$ corresponds to the start of the exponential growth and $t_3$ marks the end of the exponential growth. The simultaneous growth of the perturbed density and velocity components in the plane is expected for unstable density waves, in particular for bars. However, the vertical velocity component $U_{z,2}$ does not respond to the bar formation, and we exclude it from our analysis. 

The calculation of components of the inertia ellipsoid can also be used for the analysis of the bar growth, and to obtain the pattern speed of the bar. Slopes of the bar amplitudes and the bar strength, 
\begin{equation}
B(t) = 1 - I_{yy}/I_{xx}\ ,
\label{eq:barstr}
\end{equation}
give estimates for the growth rates, which appear to be very close to one another (and $\simeq 3$ Gyr$^{-1}$ as predicted from the rigid halo analysis).

The pattern speed is obtained from an angle of the rotation of the ellipsoid's main axes (e.g. \citealt{WPT16}). In panel (b) of Fig.~\ref{fig:bs_live} we plotted the pattern speed of the bar $\Omega_\textrm{p}$ determined by an average centred finite difference over two time-intervals of $\pm 10$~Myr before and after the evaluated time (dots) and the pattern speed after filtering out the highest frequencies. No definite structure for the pattern speed value can be found before the jump, $t<t_1$. Just after the jump, the centred finite difference shows much less scatter, and the filtered $\Omega_\textrm{p}$ varies only slowly until $t=1.9$~Gyr. Due to a bar slowdown, the pattern speed gradually decreases to 28.6~km\,s$^{-1}$kpc$^{-1}$ at $t=4$~Gyr. 

Fitting ellipses to the isophotes is one of the possibilities to quantify the bar shape~\citep{Inma}. We define a bar radius, $R_\textrm{b}$, as a radius at which the ellipticity $\varepsilon (r)\equiv 1-b_\textrm{e}/a_\textrm{e}$ ($a_\textrm{e}$ and $b_\textrm{e}$ are the major and minor semi-axes of ellipses) declines by $\sim15$\% from its maximal value. The obtained radii and ellipticities are given in Tab.~\ref{tab_rb_live}. For $t=1.0$ Gyr, the bar radius is approximately 1.1~kpc, and $\varepsilon \simeq 0.13$. At the end of the exponential growth, $t\approx 2.5$, the bar radius is 4.1~kpc, with an ellipticity of $\varepsilon \simeq 0.65$. 

\begin{table}
\begin{center}
\begin{tabular}{l l l l l l l}
\hline
Time [Gyr]             & 1.0  & 1.5 & 2.0  & 2.5 &  3.0 & 4.0 \\
\hline
$R_\textrm{b} $ [kpc]  & 1.1  & 1.1  & 2.4  & 4.1  & 5.0  & 5.7  \\
$\Omega_\textrm{p} $ [km/s/kpc]  & 48.8  & 48.0  & 45.3  & 37.8  & 33.3  & 28.6  \\
$R_\textrm{C} $ [kpc]  & 4.4  & 4.5  & 4.8  & 5.9  & 6.7  & 7.8  \\
$\varepsilon $         & 0.13 & 0.42 & 0.69 & 0.65 & 0.63 & 0.62   \\
\hline
\end{tabular}
\end{center}
\vspace{-2mm} \caption{Parameters of the bar (bar radius, pattern speed, corotation radius, ellipticity) in B-0 run at different moments in time.} 
\label{tab_rb_live}
\end{table}

With the determined $\Omega_\textrm{p} \approx 49$~km\,s$^{-1}$\,kpc$^{-1}$ at $T=1$\,Gyr, the corotation resonance occurs at $R_\textrm{C} \approx 4.4$~kpc. The ILR and OLR are located at 0.7 and 7.9~kpc, respectively. As the bar grows between $t_2$ and $t_3$, the ratio ${\cal R} \equiv R_\textrm{C}/R_\textrm{b}$ attains the value of 1.4 and remains roughly constant until the end of the simulation despite the significant slow-down of the bar in the saturation phase.

As can be seen from Fig.\,\ref{fig:bs_live}a, the bar instability in the real N-body finite thickness disc does not manifest itself as an exactly exponential amplitude growth from the early beginning of the simulation. For a long time, the amplitude is on the noise level. The start of a mild growth is possible only after a sufficiently strong wave appears in the disc  at some moment $t=t_1$, and provokes a delayed perturbation in the centre (see also top panels in Fig.\,\ref{fig:SD}). The emergence of such a wave perturbation is a random event, the probability of which is smaller for less unstable discs \citep{P16}. 

The delayed wave is a bar-like perturbation which rotates with a well-defined pattern speed (Fig.\,\ref{fig:bs_live}b), suggesting that a seed bar is already formed at $t_1$. However, the amplitude growth now is irregular and slower than exponential, until the wave amplitude $A_2$ reaches 1 or 2 per cent of the axisymmetric background $A_0$ ($t=t_2$). The exponential growth with a growth rate $\omega_\textrm{I} \approx 3$~Gyr$^{-1}$ lasts until $t=t_3$, when the instability saturates.

\subsection{Simulations with satellites}
Possible effects from satellite interactions were studied through five runs, B-1, B-1m, B-4, B-4m and B-23. The numbers in the run codes reflect number of the satellites, while `m' reflects the runs in which only the mass of the satellites was rescaled.  

\begin{figure}
\centering
   \centerline{\includegraphics [width = \linewidth]{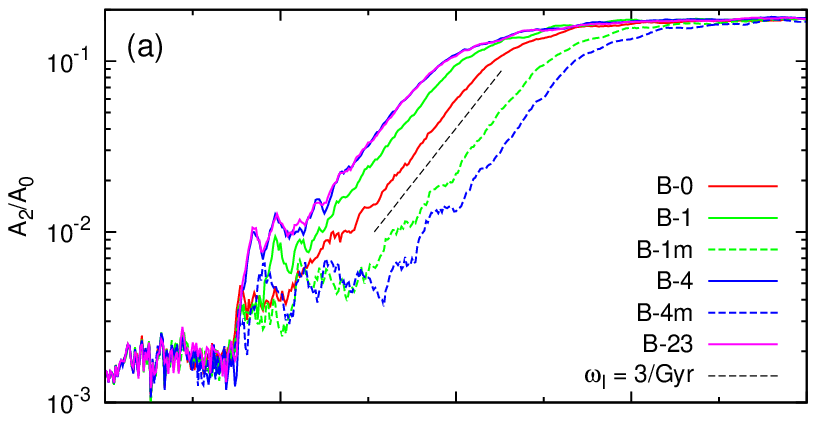}}
   \vspace{-7mm}
      \centerline{\includegraphics [width = \linewidth]{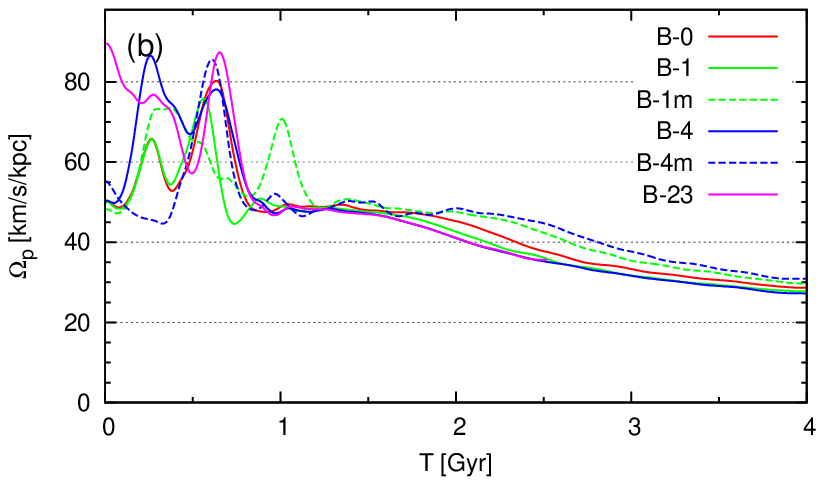}}     
   \caption{Bar formation in models with satellites: (a) bar amplitudes $A_2/A_0$ are given for B-0, B-1, B-1m, B-4, B-4m and B-23 run. The black dashed line shows the slope corresponding to a growth rate $\omega_\textrm{I} = 3$~Gyr$^{-1}$; (b) the spline smoothed pattern speeds for the same models.}
   \label{fig:ba_mass}
\end{figure}

In Fig.~\ref{fig:ba_mass} we compare the bar amplitudes $A_2/A_0$ of these models with the one calculated for B-0 (solid red). The jumps in all curves occur at the same time, $t_1 = 0.75$~Gyr, meaning that it is independent of the satellites, but rather intrinsic to the disc itself. The subsequent behaviour is different: mass rescaled runs B-1m and B-4m show a delay in bar formation, while in the fully-rescaled runs, B-1, B-4, and B-23, the bar forms earlier compared to B-0. The time difference in the bar formation $\Delta_B$ is given in Tab.~\ref{tab:runs}.

In the mass-rescaled runs, the initial positions of the satellites are further away from the host galaxy than in the fully-rescaled runs, so the interaction with the disc peaks later in the mass rescaled runs. In particular, the minimum distance in the B-1m run is at 0.96~Gyr, while in the B-1 run it occurs at 0.85~Gyr. The delay $\Delta_B$ is larger in the B-4m run where four satellites are used. 

The fully-rescaled models show contrary effect on the bar formation, compared to the mass-rescaled simulations. In the B-1 run the satellite advances the bar formation by 200~Myr (negative delay), and the higher the number of satellites, the larger the impact. In the B-4 run where the four largest satellites are used, the time difference is 300~Myr. A comparison with the B-23 run shows that the most massive satellites determine mainly the time difference -- it is only 10~Myr larger than in the B-4 run. 

In section~\ref{sec:phase_relation} we investigate in more details the true character of the observed delay/advancement.

\section{Phase relation between bar and satellite perturbations}
\label{sec:phase_relation}
A conjecture we wish to explore is focused on the measured mutual phase of an initial bar and perturbations induced by the massive satellites, and the importance towards advancement or delay in bar formation. Within the course of this section we inspect phases calculated using the surface density and velocity perturbations, assuming a simplified description of cold fluid discs.

\subsection{Perturbations from a satellite}
\label{sec:phase}
In this subsection we concentrate on the velocity perturbations in a cold fluid disc as a result of a satellite passage. In our runs, the typical encounter speed is $V \approx 400$~km\,s$^{-1}$; for a galaxy of size $d\sim20$~kpc the encounter lasts $d/V \approx 50$~Myr, which is a fraction of the rotation period of the bar, $\approx 130$ Myr, measured at the beginning of its formation.

To obtain velocity perturbations, one can integrate the acceleration of fluid particles moving on exactly circular orbits:
\begin{equation}
\d \vv(R, \theta) = -G \frac{M_\textrm{s}(t) (\vR-\vr_\textrm{s} )}{|\vr_\textrm{s} - \vR|^3} \d t\ , \quad \d\theta = \Omega(R) \d t\ ,
\label{eq:acl_sat}
\end{equation}
where $M_\textrm{s}(t)$ and $\vr_\textrm{s}(t)$ are known functions for the satellite mass and its distance from the disc centre, $\vR$ is the radius vector of a fluid particle. We take only the in-plane acceleration into account and disregard vertical perturbations here.

\begin{figure*}
\centering
   \centerline{\includegraphics [width = 150mm]{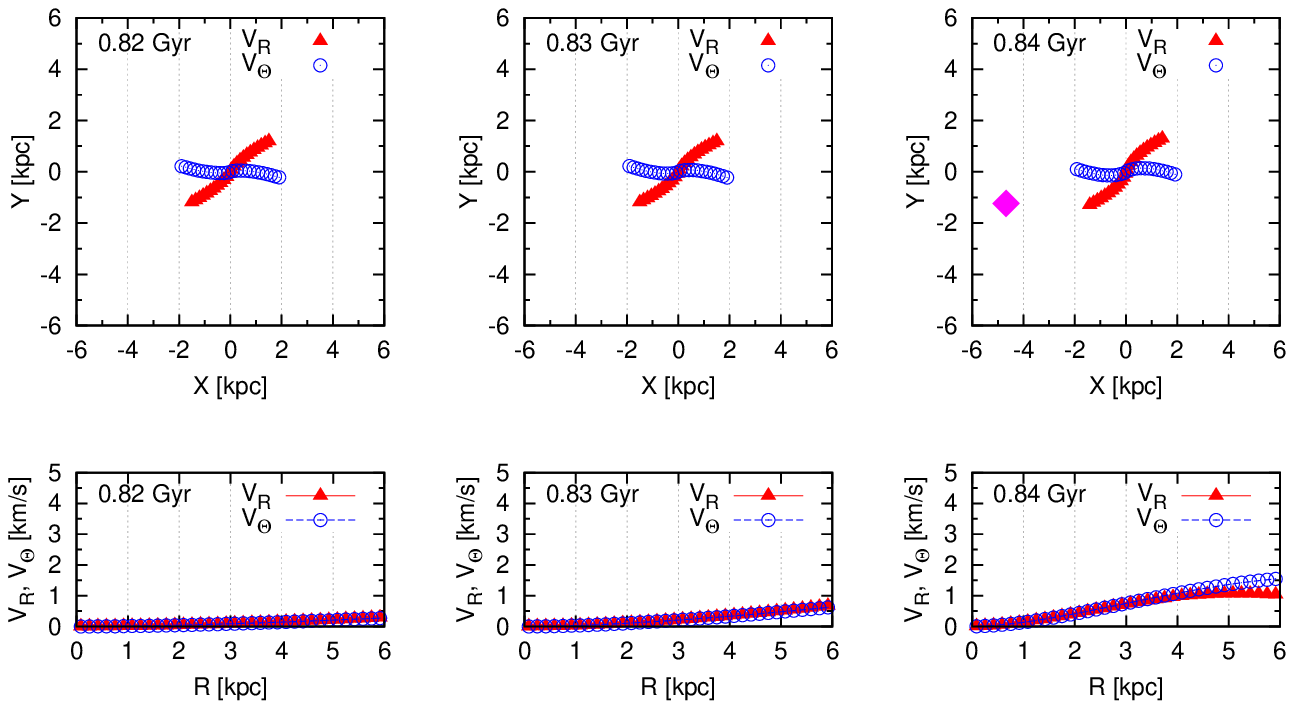} }
\vspace{-1mm}
   \centerline{\includegraphics [width = 150mm]{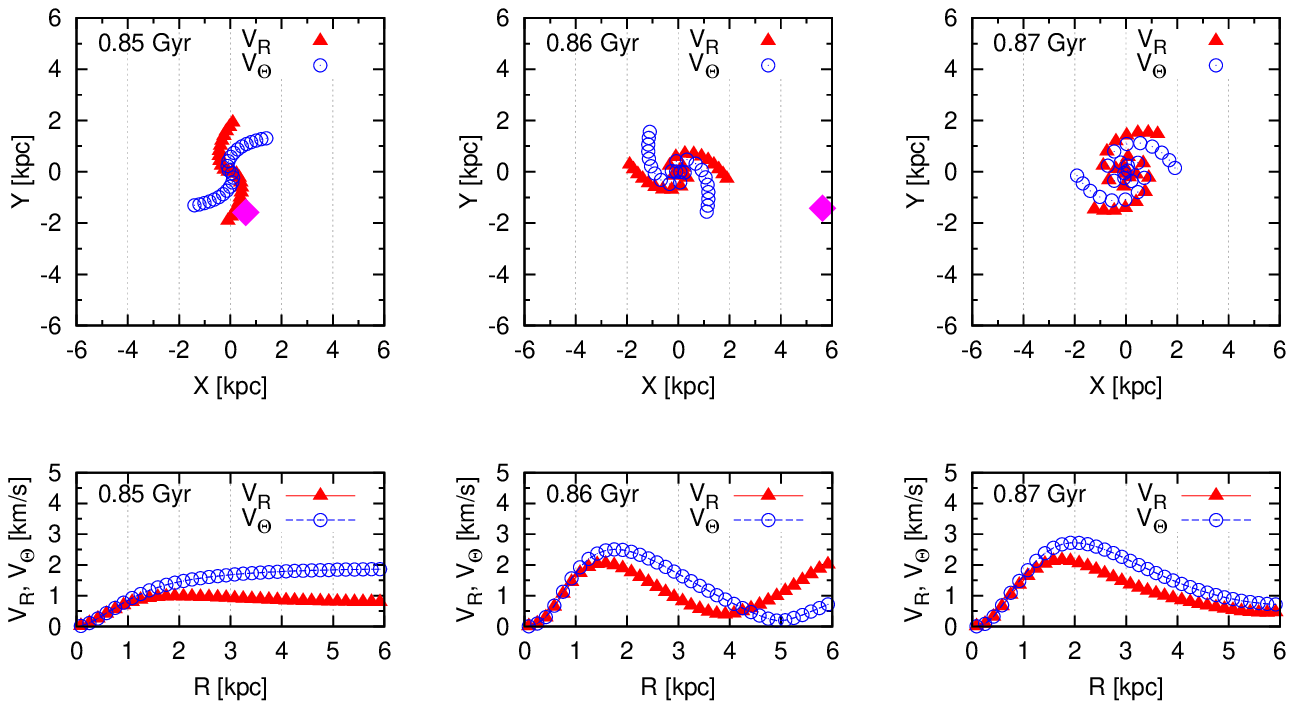} }
   \caption{Time sequence of phase (upper panels) and amplitudes (lower panels) of the $m=2$ perturbations of radial (red triangles) and azimuthal (blue circles) velocities induced by the primary satellite (pink diamonds, visible for $T=0.84 - 0.86$\,Gyr) in the cold fluid disc.}
   \label{fig:sat_v}
\end{figure*}

The phase patterns, which are irregular at the beginning of the integration, become steady and regular as the satellite approaches the disc. In Fig.~\ref{fig:sat_v}, we show amplitudes and phases of the bisymmetric $m=2$ radial and azimuthal velocity perturbations during the encounter of the primary satellite for the period between 0.82 and 0.87~Gyr. Loci of the maxima of $V_{R,2}$ and $V_{\theta,2}$ are practically unchanged from $t=0.7$ to 0.84~Gyr. After the fly-by, the maxima lines begin to wind up, and eventually turn into a tightly wound spiral. The amplitudes of velocity components grow up to $t=0.86$~Gyr, then remain practically unchanged. 

\subsection{Phase relations in the cold fluid disc}
Equations of fluid dynamics are employed frequently to analyse phenomena taking place in stellar discs \citep[e.g.][]{Ber14}. Indeed, when the stellar disc is cold enough, mean radial and azimuthal velocities can be obtained from linearised Euler equations for zero-pressure discs, because a cold stellar disc is dynamically equivalent to a fluid disc with zero pressure. If the perturbation has $m$-fold rotation symmetry and $\propto \textrm{e}^{i(m\theta-\omega t)}$, the perturbed surface density $\Sigma_{da}$ and average velocity components ($v_{\theta a}, v_{R a}$) obey the following relations \citep{BT08}:
\begin{equation}
v_{Ra}(R) = \frac{i}{\Delta} \Big[ (\omega-m\Omega) \frac{\d\Phi_a}{\d R} - \frac{2m\Omega\Phi_a}{R}  \Big]\ , 
\label{eq:vr}
\end{equation}
\begin{equation}
v_{\theta a}(R) = -\frac{1}{\Delta} \Big[ 2B \frac{\d\Phi_a}{\d R}  + \frac{m(\omega-m\Omega)\Phi_a}{R}  \Big]\ , 
\label{eq:vp}
\end{equation}
\begin{equation}
-i(\omega-m\Omega) \Sigma_{da} + \frac{1}R \frac{d}{dR}(Rv_{Ra}\Sigma_0) + \frac{im\Sigma_0}{R} v_{\theta a} = 0\ , 
\label{eq:sa}
\end{equation}
where $B \equiv -\kappa^2/4\Omega$ is the Oort constant; $\Sigma_0$ is the disc surface density and $\Phi_a$ is the potential perturbation;
\begin{equation}
\Delta \equiv \kappa^2 - (\omega-m\Omega)^2
\label{eq:delta}
\end{equation}
is positive between the Lindblad resonances.

In this analysis, we shall distinguish two extreme cases: tightly wound spirals and the bar. In the {\it tightly wound} approximation, a perturbed quantity (e.g. the potential) can be written in the form   
\begin{equation}
\Phi_a(R) = |\Phi_a| \textrm{e}^{iF(R)} = |\Phi_a|  \textrm{e}^{i\int^R k\, dR} \ , 
\label{eq:pot}
\end{equation}
with phase function $F(R)$, where $k=dF(R)/dR$ and $|kR| \gg 1$. From eq. \ref{eq:sa} one has
\begin{equation}
kv_{Ra}\Sigma_0 = (\omega-m\Omega) \Sigma_{da}\ , 
\label{eq:sa1}
\end{equation}
that is, surface density of the wave is in phase with the radial velocity outside corotation, and in anti-phase inside corotation:
\begin{equation}
\begin{array}{rcl} 
        F_{R} - F_{\Sigma}  = \pi & : & R<R_C \ ,\\ 
        F_{R} - F_{\Sigma}  = 0 & : & R>R_C \ .
\end{array}
        \label{eq:F1}
\end{equation}

The equation for the azimuthal velocity (eq. \ref{eq:vp}) simplifies to
\begin{equation}
v_{\theta a}\Sigma_0 = \frac{-2iB}{\Delta} k \Phi_a \ , 
\label{eq:vp1}
\end{equation}
and taking into account that $\Phi_a = -2\pi G\Sigma_{da}/|k|$,
\begin{equation}
F_{\theta} - F_{\Sigma} = -\pi/2\ , 
\label{eq:F2}
\end{equation}
meaning that the surface density lags behind the azimuthal velocity by $\pi/4$ for an $m=2$ perturbation.

In the {\it bar region} we assume that the radial derivative is negligible compared to the other term in square brackets in (\ref{eq:vr}, \ref{eq:vp}). For the radial velocity,
\begin{equation}
v_{Ra}(R) = -\frac{2im}{\Delta} \frac{\Omega}{R} \Phi_a  \ ,
\label{eq:v2}
\end{equation}
that is, 
\begin{equation}
F_{R} - F_{\Sigma} = \pi/2\ , 
\label{eq:F3}
\end{equation}
meaning that radial velocity lags behind the surface density by $\pi/4$ ($m=2$). For the azimuthal velocity,
\begin{equation}
v_{\theta a}(R) = -\frac{m}{\Delta}  \frac{(\omega-m\Omega)}{R} \Phi_a  \ ,
\label{eq:vp2}
\end{equation}
that is, 
\begin{equation}
\begin{array}{rcl} 
        F_{\theta} - F_{\Sigma} = \pi & : & R<R_C \ ,\\ 
        F_{\theta} - F_{\Sigma} = 0 & : & R>R_C \ .
\end{array}
        \label{eq:F4}
\end{equation}

\subsection{Phase relations in the stellar discs}
To compare the phases of surface density and velocity perturbations in the stellar discs, we calculate a relative surface density, $A_2(R,t)/A_0(R,t)$,
\begin{equation}
A_m(R,t) = {\sum\limits_{j}}' \mu_j \textrm{e}^{-im\theta_j}\ , 
\label{eq:ART}
\end{equation}
and velocity components, $V_{k,2}(R,t)/A_0(R,t)$, 
\begin{equation}
V_{k,m}(R,t) = {\sum\limits_{j}}' \mu_j \tilde v_{k,j} \textrm{e}^{-im\theta_j}.
\label{eq:VRT}
\end{equation}
In both expressions, the summations are taken over disc particles within a ring $\Delta R$ near $R$. 

\begin{figure*}
\centering
   \centerline{\includegraphics [width = 150mm]{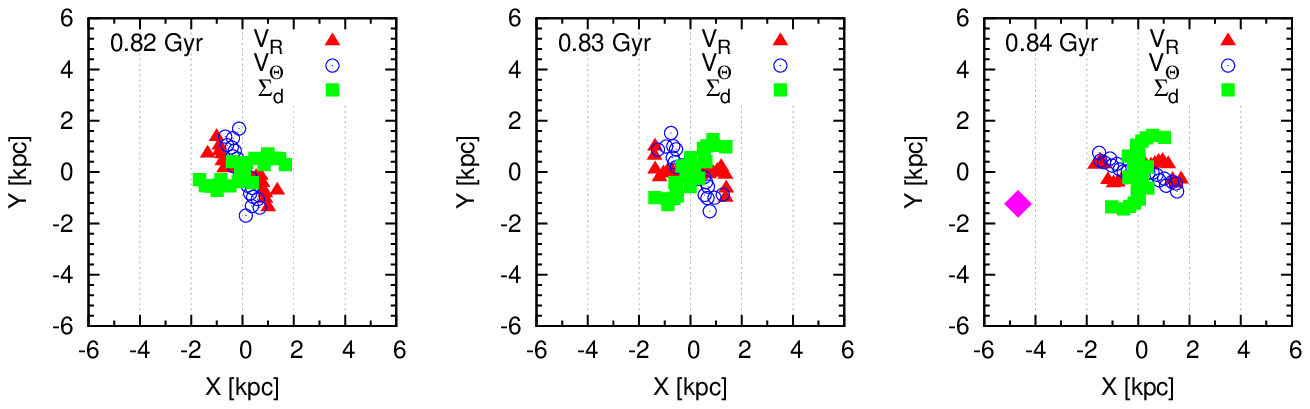}}
\vspace{-1mm}
   \centerline{\includegraphics [width = 150mm]{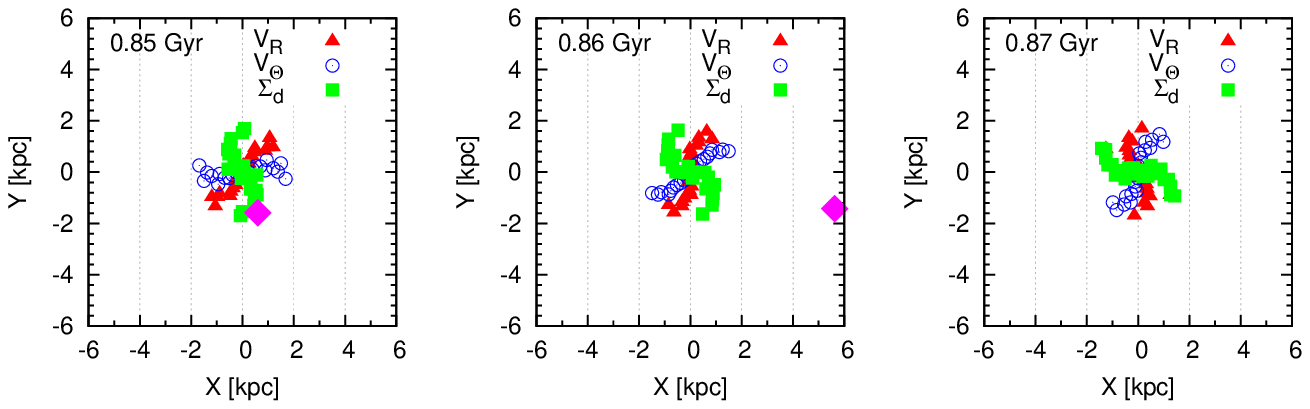}}
   \caption{Loci of maxima of $m=2$ perturbations (velocity and surface density) in the B-0 run (no satellites) at the time of the encounter in B-1 (the pink diamond shows the position of the primary satellite in B-1 for visualisation).}
   \label{fig:b0_v}
\end{figure*}

Fig.~\ref{fig:b0_v} shows the loci of maxima $\theta(R) = -F/m$ for the B-0 run without satellites. Note that the maxima of $V_{\theta,2}$ are shifted by $\pi/2$ from the surface density maxima, and the maxima of $V_{R,2}$ lag by $\pi/4$ from the surface density maxima, in accordance with the phase relations for a bar in the cold fluid disc derived above. 

The comparison with Fig.\,\ref{fig:sat_v} shows that the wave velocity components are approximately in phase with disc perturbations caused by a satellite when the satellite approaches the disc centre, but become out of phase after $t=0.85$~Gyr.  

\begin{figure*}
\centering
   \centerline{\includegraphics [width = 150mm]{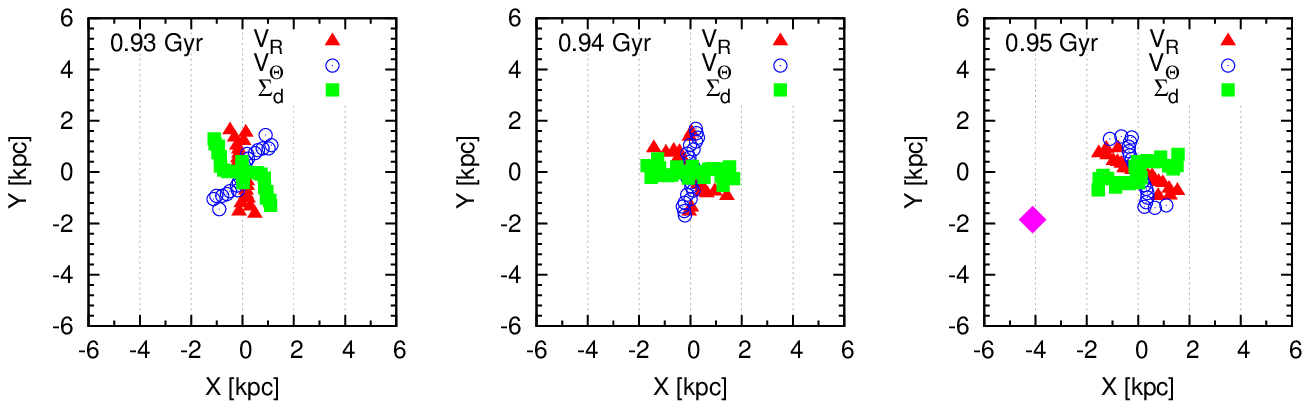} }
\vspace{-1mm}
   \centerline{\includegraphics [width = 150mm]{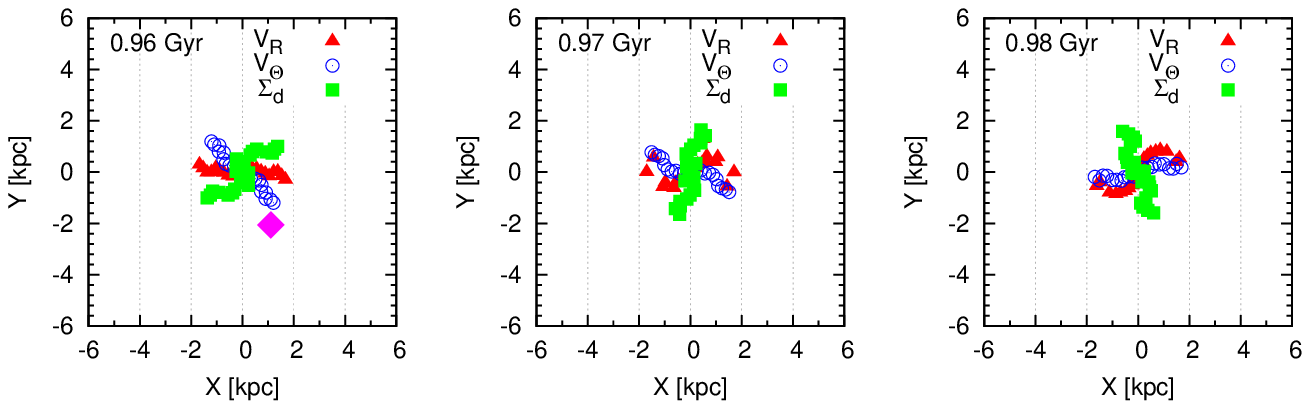} }
   \caption{Same as in Fig.\,\ref{fig:b0_v}, but at the time of the encounter in B-1m.}
   \label{fig:b1m_v}
\end{figure*}

Fig.~\ref{fig:b1m_v} is similar to Fig.~\ref{fig:b0_v}, but for the primary satellite in the mass scaled models. The orientation of the bar is roughly perpendicular to that of the bar in Fig.~\ref{fig:b0_v} at the corresponding moments of time (time difference is $\sim 110$~Myr). So in this case the bar is out of phase when the satellite approaches the pericentre. 

\begin{figure}
   \centering
   \includegraphics [width = \linewidth]{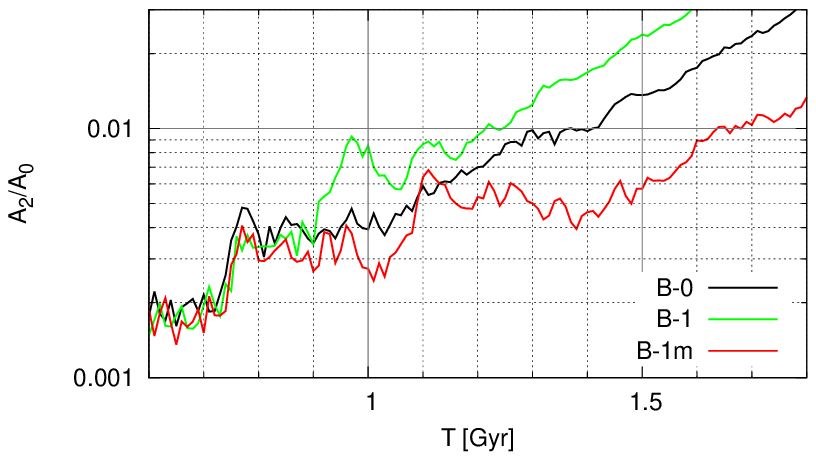}
   \caption{Bar amplitudes $|A_2/A_0|$ for the B-0, B-1, B-1m runs (selected zoom of Fig.~\ref{fig:ba_mass}).}
   \label{fig:tsd_3}
\end{figure}

The derived difference in phases during the satellite approach is essential, but seems to be only part of a more complex story. It follows from Fig.~\ref{fig:ba_mass}, where one can see a rather complicated bar amplitudes' behaviour. In Fig.~\ref{fig:tsd_3} we show a zoom-in of the early growth phase for the B-0, B-1, and B-1m runs. In particular, we stress the following features. 

In the fully-rescaled B-1 run, the maximum occurs at $t=0.97$ Gyr, that is, 120~Myr (almost one period of bar rotation) after the pericentre passage; the 0.97~Gyr maximum is followed by minima at 1.06~Gyr and 1.16~Gyr. In the mass-rescaled model B-1m, a deep minimum occurs at $t=1.01$~Gyr, that is, 50~Myr after the pericentre passage; the minimum is followed by a large maximum exceeding the amplitude in the B-0 run at 1.11~Gyr followed by a nearly constant behaviour until 1.5~Gyr. These features cannot be understood from simple phase arguments as discussed above. 

\subsection{Interference of bar and satellite perturbations}
\begin{figure*}
   \includegraphics [width = 85mm]{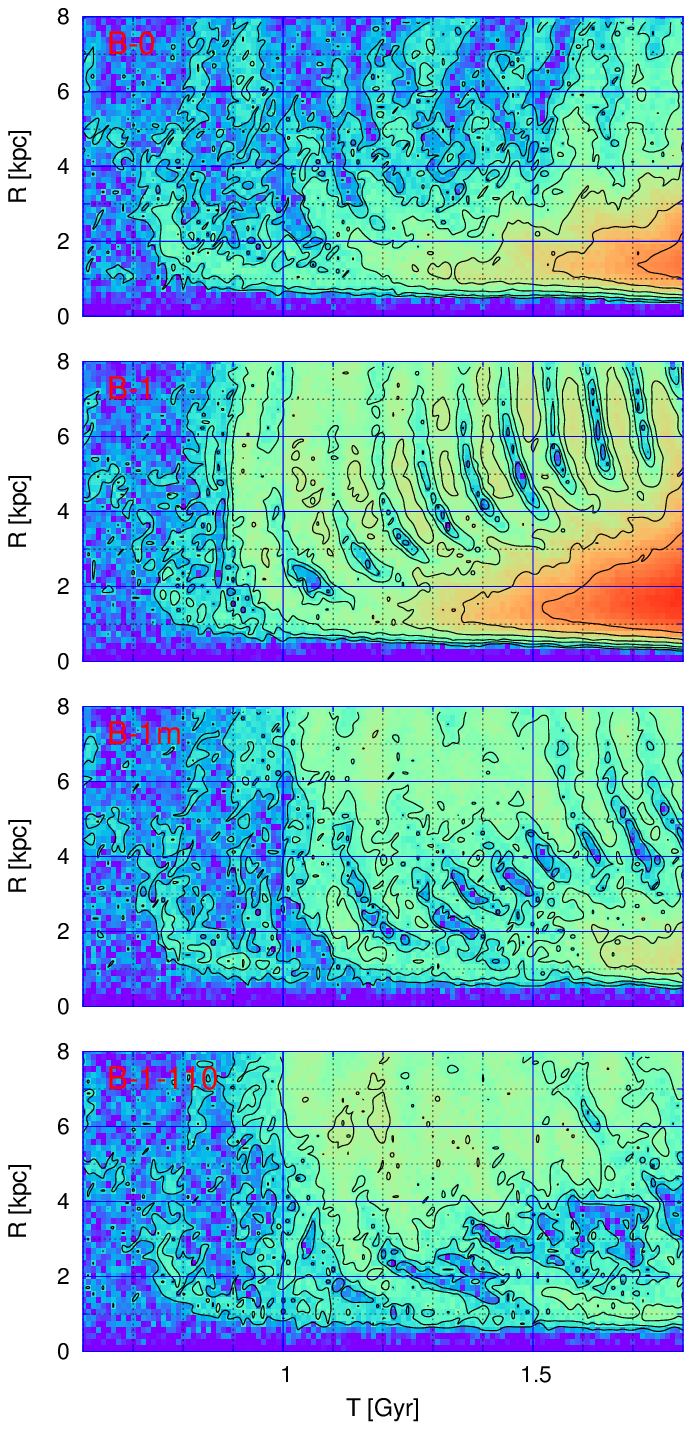} \hspace{1mm} \includegraphics [width = 85mm]{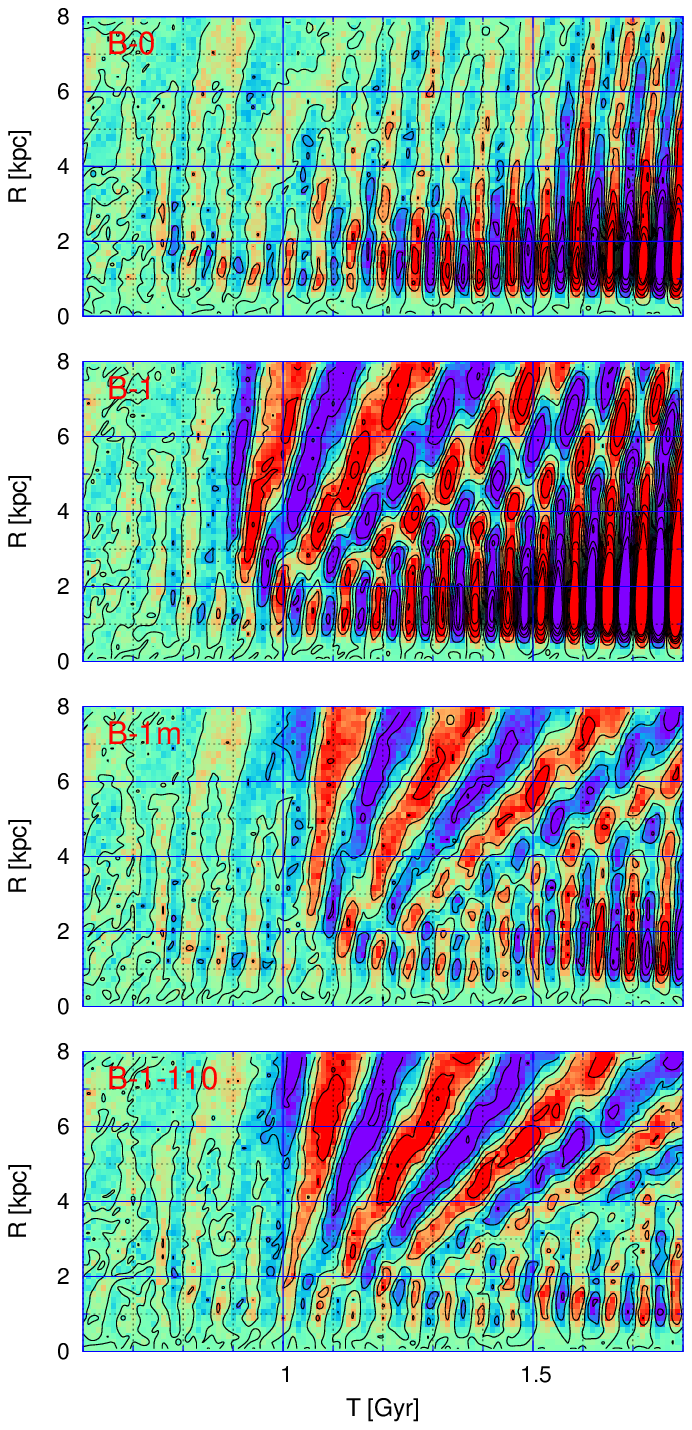}
   \caption{Surface density maps  $\log|A_2/A_0|$ (left panels, red = high, blue = low amplitude); $\Re A_2/A_0$ (right panels, red = positive, blue = negative amplitude at the $x$-axis) for B-0, B-1, B-1m, and B-1-110 runs (top to bottom), reflecting the interference between the bar-like perturbation in the centre, random perturbations, and tidally induced waves. The time period is $0.6<t<1.8$ Gyr.}
   \label{fig:SD}
\end{figure*}

The interaction between the central bar-mode and a perturbation from the satellite can be studied using density surface maps showing the total bisymmetric ($m=2$) amplitude in the $(R, t)$-plane. Left panels in Fig.\,\ref{fig:SD} show $\log|A_2/A_0|$ for the isolated run B-0, as well as for one satellite runs B-1 (fully-rescaled) and B-1m (mass-rescaled). The lower left panel contains the map for not-yet-mentioned B-1-110 run, which differs from B-1 by a 110~Myr time delay of the satellite encounter. The lag is calculated to mimic a passage of the satellite near the pericentre of the mass-rescaled B-1m run (see Fig.\,\ref{fig:sat_orbit_bs}).  

The descending `tongues' observed in the panels are nothing but interference patterns of two waves. Even in the upper figure for the isolated B-0 run, it is the interference between a small bar-like perturbation mostly residing in the central part, and random perturbations of the disc at $R \sim 1...2\,R_\textrm{d}$. These perturbations may act in accordance and accelerate bar formation, or, on the contrary, destroy one another.

The random peripheral perturbations are repeated in the lower panels. In addition, they contain imprints from the satellite for the different cases B-1, B-1m, B-1-110.
In Fig.\,\ref{fig:b1} one can see for case B-1 the onset of the disc perturbation after the encounter (0.87 and 0.89 Gyr), which clearly has a bar-like global shape and obeys phase relations obtained for bars in Section 5.2. This disc perturbation evolves so that its shape winds up gradually to the tightly wound spiral (last row of Fig.\,\ref{fig:b1}; the subsequent evolution is not shown here), but also it causes a disc response in the central region similar to those presented in T81; we refer to Figs. 1, 2 therein. Recall that Toomre analysed responses from weak corotating (Fig.1) and immobile (Fig.2) transient imposed dipole forces in a {stable} Mestel disc. The disc shows strong spiral responses developing from the centre outward, and sheared with a speed comparable with material arm shear, $\Omega(R)$. This is a manifestation of `a strong cooperative effect' called {\it swing amplification}. For this mechanism to be efficient, the parameter 
\begin{equation}
X\equiv \frac{\kappa^2 R}{2m \pi G \Sigma_0}
\label{eq:Xt}
\end{equation}
should not exceed 3, which is the case in our model in the range $0.6<R<5.5$~kpc, with a minimum $X\simeq 2$ at $R\simeq 2.3$~kpc.

Much like in Toomre simulations, we expect similar responses to the disc external perturbations imposed by the satellite. Such responses grow on the time-scale of one period in the centre and interact with a small bar, strengthening or weakening the latter. 

Right panels of Fig.~\ref{fig:SD} show $\Re A_2/A_0$, that is, a joint $m=2$-amplitude along the $x$-axis, for the same runs B-0, B-1, B-1m, B-1-110 from top to bottom. In the upper panel, one can see red and blue spots standing for maxima and minima and showing emergence of the bar, starting from $t\sim 0.75$~Gyr ($R<2$~kpc). These spots extend vertically in time as the bar length grows. Also one can note smaller slightly shifted spots above the bars. These are interference patterns between the shearing external disc response to the encounter and the spiral arm adjacent to the rotating bar in the centre. 

\begin{figure}
  \centering
   \includegraphics [width = \linewidth]{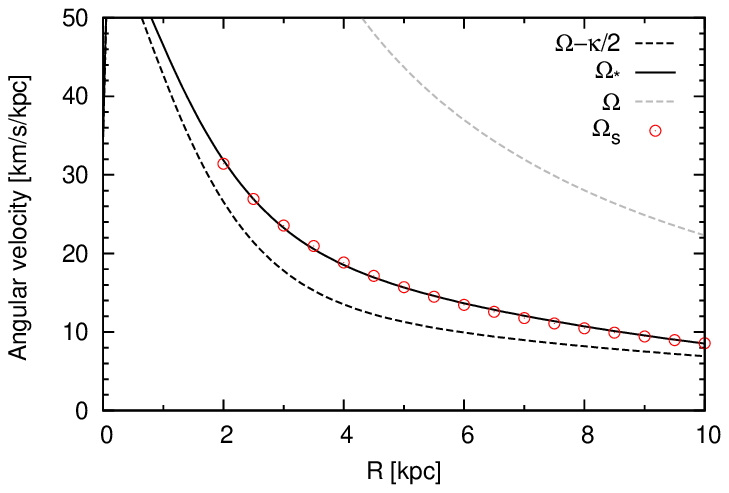}
   \caption{Local pattern speeds $\Omega_s$ of the wave patterns (red circles) compared to main frequencies of the disc ($\Omega$ is the angular velocity, $\kappa$ is the epicyclic frequency, $\Omega_*$ is given by (\ref{eq:Oms})).}
   \label{fig:wkbw}
\end{figure}

The interference spots are localised roughly along the straight lines with inclination growing in time. Using these lines one can infer periods of the interference pattern as a function of radius, which roughly obey the linear law, $T_\textrm{i} = aR+b$. Since frequency difference between the bar and the spiral response $\Omega_\textrm{b}-\Omega_\textrm{s}(R) = 2\pi/T_\textrm{i}$, one can calculate the local pattern speed $\Omega_\textrm{s}(R)$ for wave patterns due to the encounter. This is shown by red circles in Fig.~\ref{fig:wkbw}, along with $\Omega$ and $\Omega-\kappa/2$ curves.

Local pattern speeds $\Omega_\textrm{s}$ of the wave patterns can be explained using the simplest fluid WKB dispersion relation. Each wave pattern evolves according to the relation, 
\begin{equation}
\omega = \omega(k,R),
\label{eq:WKB_g}
\end{equation}
meaning that $R = R(k,\omega)$ is a function of $k$; $\omega \equiv m\Omega_\textrm{s}$ is a parameter. For very open spiral\footnote{Application of WKB theory should be treated here as extrapolation only.}, $k$ is close to zero, so 
$\Omega_\textrm{s} = \Omega(R) - \kappa(R)/m$, and $R$ is close to the ILR, $R_\textrm{ILR}$. As the local wave pattern evolves, $(\omega-m\Omega)^2$ attains its minimum at some $k=k_*$, so $R$ attains its maximum \citep{T69}. Then $k_*$ can be estimated from the dispersion relation for fluid discs:  
\begin{equation}
(\omega - m\Omega(R))^2 = \kappa^2(R) - 2\pi G \Sigma_0 |k| + k^2 c^2\ ,
\label{eq:WKB}
\end{equation}
where $k$ is the wavenumber and $c$ is a sound speed, which we use interchangeably with the radial velocity dispersion. Thus
\begin{equation}
k_* = \frac{\pi G \Sigma_0}{c^2}\ ,
\label{eq:k_s}
\end{equation}
and substitution to the WKB relation gives
\begin{equation}
\Omega_* (R) = \Omega(R) - \frac{\kappa(R)}m (1-Q(R)^{-2})^{1/2}\ .
\label{eq:Oms}
\end{equation}
In the last expression, $Q$ denotes Toomre stability parameter for fluid discs,
\begin{equation}
Q(R) = \frac{\kappa(R)c(R)}{\pi G \Sigma_0(R)}\ .
\label{eq:Qt}
\end{equation}

As is seen from Fig.~\ref{fig:wkbw}, $\Omega_*$ and $\Omega_\textrm{s}$ agree relatively well, which proves that the picture presented here is an interference pattern between the bar and the density wave. We note that due to large $Q$ ($>2$), radial excursions of the wave patterns are small, and are localised close to their ILR.

The interference of the bar and the delayed response to the disc perturbation in the B-1 run leads to a bar enhancement. This is seen from comparison of the right panels in the first and second rows of Fig.~\ref{fig:SD}. In particular, a red bar spot immediately after $t=1$~Gyr in the B-1 run is definitely stronger than its counterpart in the B-0 run. 

Contrary to simulations performed in T81, our external perturbation does not appear for only a short time. Thus it provides a number of delayed responses in the centre which can also either enhance or destroy the bar, unless the external perturbation becomes tightly wound. In case of B-1, it also works for enhancement. 

In the last two rows, we see the opposite behaviour. In B-1m, the red bar spot immediately after $t=1$~Gyr is definitely weaker than in B-0 run, and it is seen that the delayed response is not in phase with the original bar. Moreover, it seems that the established perturbation also suffers from the subsequent delayed response waves. 

We also note powerful random waves seen in B-0 panels between 1.05 and 1.3~Gyr, and 1.15 and 1.4~Gyr. They also came into play and decrease amplitudes in B-1m and B-1-110 runs.

\section{The early encounter}
\label{sec:early}
In the previous section we argue that satellite encounters may cause a delay or advancement in bar formation after the small bar is already formed in the centre ($t>t_1$), but its amplitude is still insignificant ($t<t_2$). Here we determine what occurs if the encounter takes place well before $t_1$. For this purpose, we performed B-1-500 run, in which the fully scaled satellite approaches the host galaxy 500~Myr earlier than in the B-1 run. 

In the surface density maps (Fig.\,\ref{fig:SD500}), we see disc perturbations due to the encounter immediately after $t=0.35$~Gyr. While there is no visible random perturbation in the disc (see upper left panel), the only perturbation that exists is the one due to the satellite and subsequent delayed response in the central region ($R<2$ kpc), visible only after $t=0.4$~Gyr. Then it disappears at 0.6~Gyr, and a second perturbation appears after a short gap. Once the first random wave has reached the centre at $t \simeq 0.75$~Gyr, the large scale spiral becomes tightly wound and ineffective in providing a delayed response in the centre. This follows from comparison of the amplitudes in the left panels for $R<2$~kpc.

\begin{figure}
   \centering
   \includegraphics [width = \linewidth]{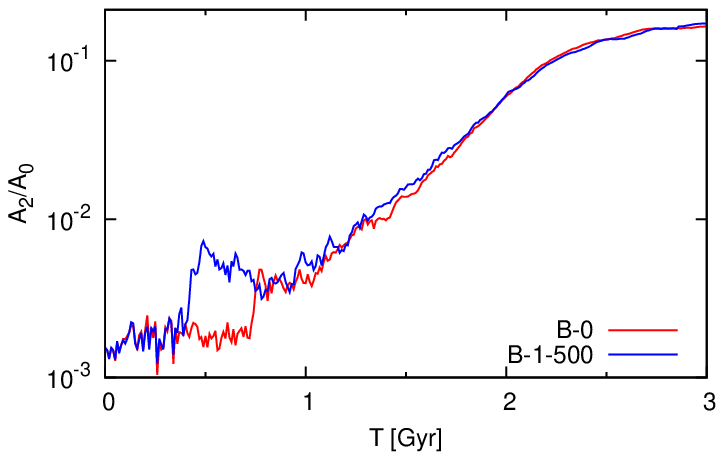}
   \caption{Bar amplitudes for B-0 and B-1-500 runs.}
   \label{fig:rtmaps}
\end{figure}

The resulting bar amplitude calculated for disc particles within characteristic scale length $R_\textrm{d}$ is shown in Fig.~\ref{fig:rtmaps} together with the bar amplitude of the B-0 run. It is seen that after a jump at $t\simeq 0.4$~Gyr due to the delayed response waves, the central perturbation gradually fades away until the random wave forms a bar perturbation at $t\simeq 0.75$ Gyr -- seen already in the previous runs. Afterwards, the B-1-500 curve continues almost in pace with the B-1 curve.

\section{Comparison with observations}
\label{sec:observations}
In this section we would like to discuss how our results compare with observations. Throughout this paper we have argued that the Aq-D2 is believed to be a fair representative of a Milky Way-like host. Fig.~3 in \citet{moetazedian} is very helpful for comparing the distribution of satellites from the Aq-D2 run (red data points) with observations of our Milky Way. According to this figure, there resides a handful of massive satellites with $M_\textrm{tid} > 10^9 M_\odot$ among the seven simulations with pericentre distances within 25~kpc from the centre. The Milky Way's Sagittarius dwarf galaxy (Sgr) has an estimated current total mass $M_\textrm{Sgr} \sim 5 \times 10^8 M_\odot$ and a pericentre of $\sim20$~kpc \citep*{law}. If we assume Sgr to be on its second passage, then it has probably lost $\sim$50\% of its mass during the initial crossing. In the Aq-D2, an analogue to Sgr can be found at $\sim20$~kpc. In addition, more massive Milky Way satellites such as SMC with a mass $\sim 6.5 \times 10^9 M_\odot$ \citep{bekki} and galactocentric distance of 60~kpc \citep{smc} have a corresponding analogue in the Aq-D2 realisation. 

As previously shown by Eq.~\ref{eq:dV}, the impact of a satellite with a given tidal mass and a pericentric distance $b$ scales as $\sim M_\textrm{tid}/b^2$. For a Sgr or SMC-like satellite we expect smaller impact as their passage occurs at a further distance to the centre (50-60~kpc) and hence, we fall below $10^{-4}$ (Fig.~\ref{fig:sat_stat}). 

The distribution of satellites from both N-body cosmological simulations and observational studies are dominated by lower mass objects. In the Aquarius satellites distribution there exists a fair number of such low-mass satellites with $M_\textrm{tid} < 10^9 M_\odot$ and pericentres within the inner 10~kpc of the disc. Observations of the Local Group have shown the existence of so-called ultra-faint dwarf galaxies with velocity dispersions of the order of 10~kms$^{-1}$ \citep{simon}. Recent observations as part of the Dark Energy Survey (DES) have increased the total number of detected ultra-faint dwarfs to 17 \citep{des}. This acts in favour of narrowing down the gap between cosmological model predictions and observations at the lower mass end of the galaxy mass function, that is, ``the missing satellite problem''. The DES year-two quick release (Y2Q1) predicts a total of $\sim$100 ultra-faint dwarfs in the full sky cover. These objects are highly dark matter dominated with a very low stellar budget. In order to estimate the DM mass content of the satellites, their velocity dispersion needs to be measured using spectroscopy; this has only been done for a few satellites (e.g. \citealt{martin16,simon16}). These galaxies would be subject to strong tidal stripping in case of very close encounters with the disc, meaning it is very difficult to observe any remnants. Therefore, our current observational sample of ultra-faint dwarfs is limited and fainter magnitudes need to be reached. 

There exists an inconsistency in current models of the Milky Way bar, where a broad range of values are discussed for bar properties such as age, pattern speed etc. (e.g.~\citealt*{wegg,monari}). Therefore we are unable to constrain our models,  making it impossible to determine the age of the Milky Way's bar.      

\section{Conclusions}
\label{sec:summary}
In this paper we analyse how the bar formation time can be affected by dark matter satellite encounters. Our Milky Way-like galaxy model is studied mainly using N-body simulations. All components, including an exponential stellar disc, a NFW halo, and a S\'{e}rsic cuspy bulge are tailored to the Milky Way and are live, that is, represented by particles that interact according to Newton's law of gravity. 

The adopted parameters of the host galaxy provide a disc mildly sensitive to bar instability with a characteristic e-folding time (inverse growth rate) $\tau_\textrm{e} = 340$~Myr. This time, as well as the bar pattern speed, $\Omega_\textrm{p} = 48...49$~km\,s$^{-1}$kpc$^{-1}$ were obtained from our simulations and agree well with global mode calculations in the framework of linear perturbation theory. Such a large e-folding time is primarily due to the relatively high radial velocity dispersion (35 km\,s$^{-1}$ in the Solar neighbourhood), and a weak central cusp with $\rho \propto r^{-0.5}$. The cusp does not prohibit bar formation, as long as a disc with finite thickness rather than a razor-thin disc is considered \citep{PBJ16a}.

Bar formation in the isolated galaxy occurs in the presence of random swing amplified wave packets appearing in a range of radii with Toomre parameters $X<3$ and $Q<3$ (between 2 and 4 kpc in our case). These packets interfere with a bar-like perturbation being a density wave in the inner part, which grows on its own due to bar instability. The interference leads randomly to demolition or enhancement of the bar. The latter is seen as a jump of the bar amplitude at some arbitrary moment in time ($t_1$ in Fig.\,\ref{fig:bs_live}). However, the exponential growth occurs later after additional random wave packets reinforce the inner bar, thus the moment $t_2$ at which the bar starts to grow exponentially is also a random quantity \citep[see also][]{P16}.

The properties and orbital parameters of the satellite galaxies are derived from the Aquarius D-2 cosmological simulation. All satellites with tidal mass $M_\textrm{tid}$ above $10^8 M_\odot$ and pericentric distances to the host galaxy below 25~kpc were selected. Our set of runs with satellites consists of four runs with one satellite, two runs with four satellites (with tidal mass M$_\textrm{tid} \geq 10^9 M_{\odot}$) and one run with all 23 satellites. Each satellite was represented by 50,000 particles. In all cases, the pattern speed and the e-folding time of the bar in the host galaxy remain unchanged. However, we noted that from run to run, the bar amplitudes grow asynchronously. In the earliest case, the bar forms 300~Myr earlier than in the reference isolated run without satellites. In the latest run, the bar appeared with a 600~Myr delay. Most of the advancement (or delay in the other two cases) is caused by the primary satellite whose pericentre is the closest to the galactic centre. The effect from other massive satellites is appreciable, yet turns out to be less significant. The lighter satellites play practically no role.

Satellites add complexity to the physical picture of bar formation. One satellite on the same orbit can give both advancement and delay, simply by passing in a slightly different time. Thus, orientation of the bar is important. However, it is difficult to specify `a rule of thumb' for the sign of the effect, since the satellite angular speed is at least two times higher than the bar pattern speed, so a range of various bar phases is seen in all cases. 

\begin{figure*}
   \centerline{\includegraphics [width = 150mm]{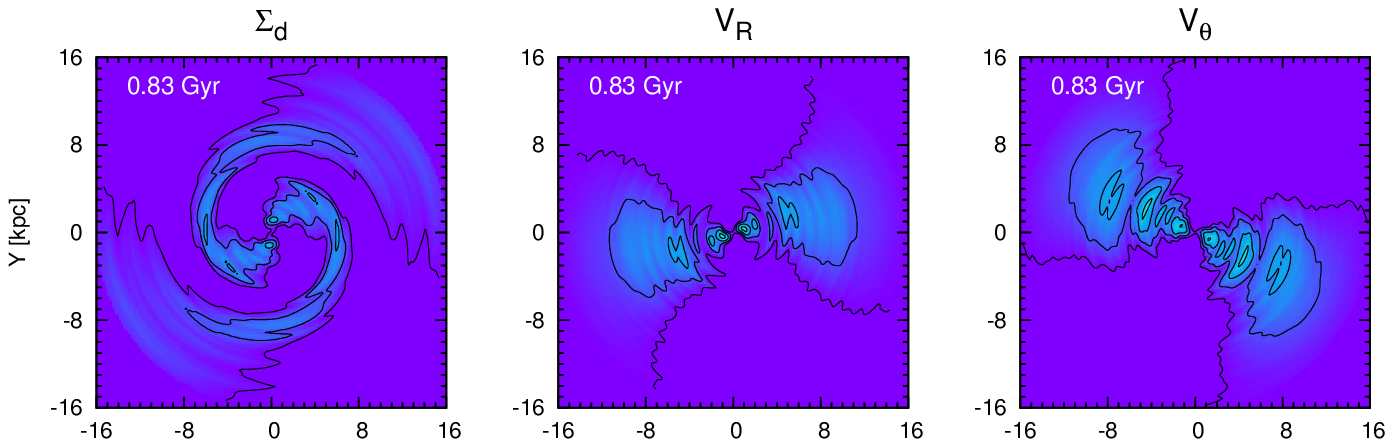} }
\vspace{-6mm}
   \centerline{\includegraphics [width = 150mm]{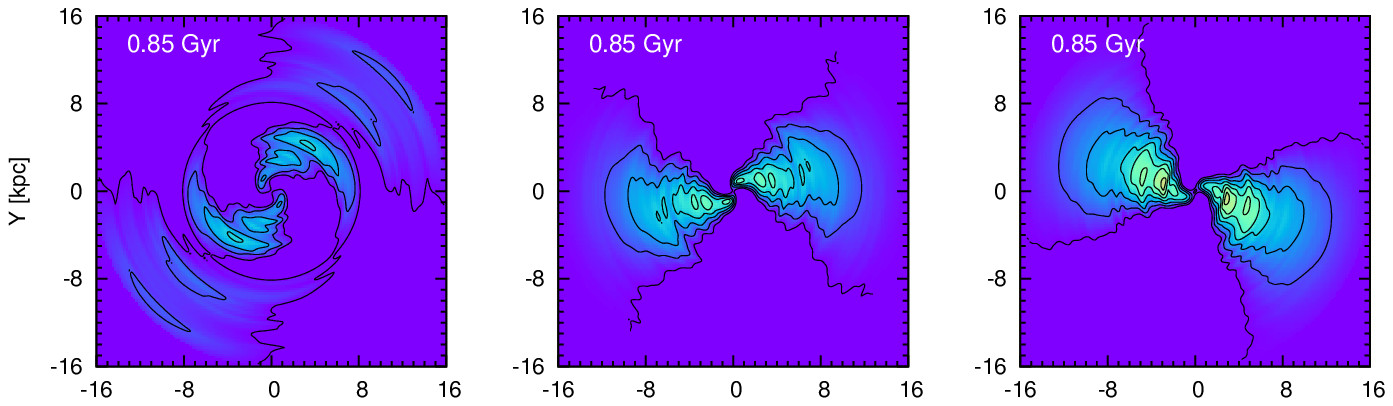} }
\vspace{-6mm}
   \centerline{\includegraphics [width = 150mm]{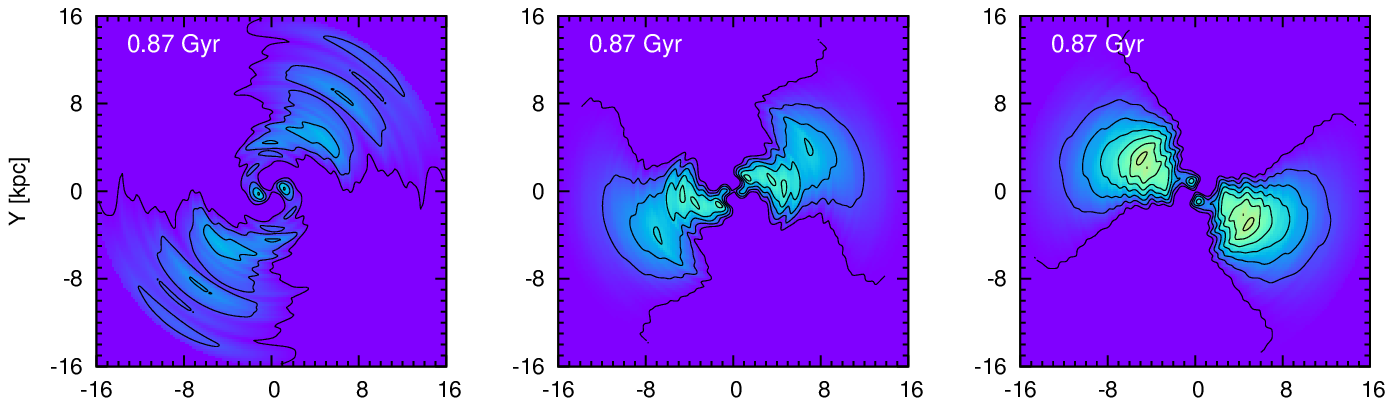} }
\vspace{-6mm}
   \centerline{\includegraphics [width = 150mm]{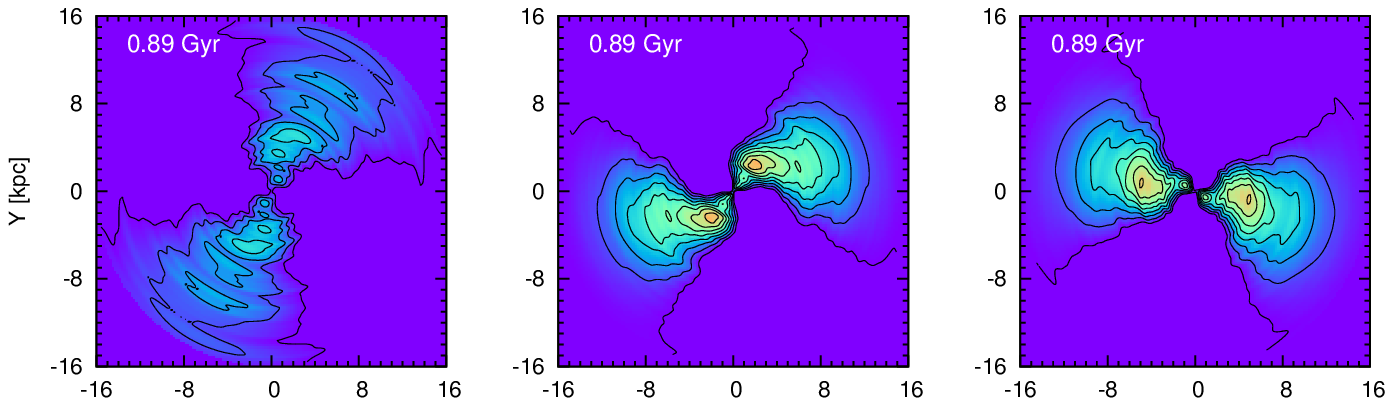} }
\vspace{-6mm}
   \centerline{\includegraphics [width = 150mm]{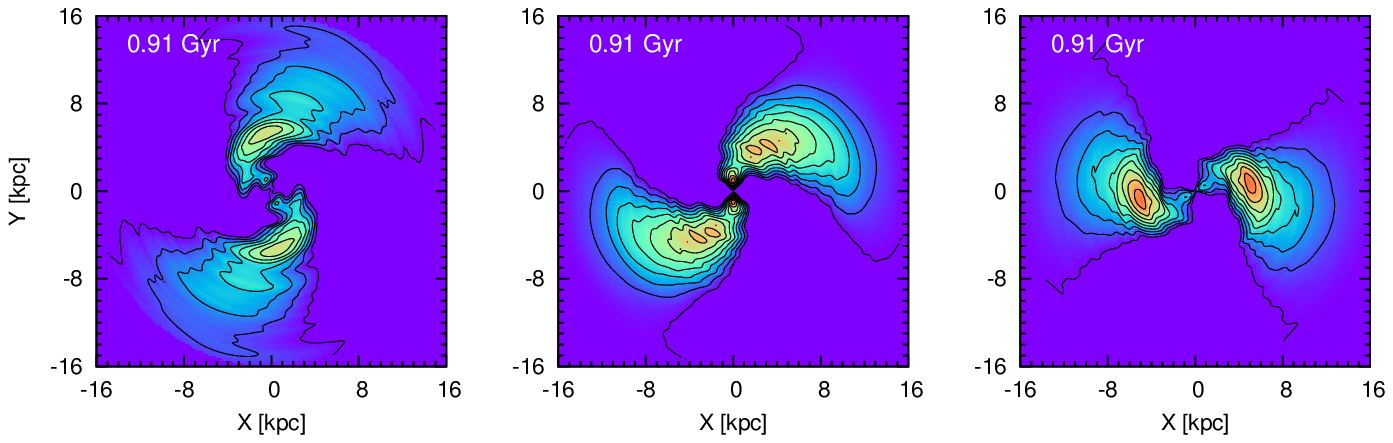} }
   \caption{Onset of the large-scale bar-like perturbation from the satellite in the B-1 run (time from top to bottom, $\Sigma_\textrm{d}$, $V_\textrm{R}$ and $V_\theta$ from left to right).}
   \label{fig:b1}
\end{figure*}

The main features of the satellite impact on the bar formation can be understood again through the wave interactions. After the satellite passage, a tidal wave is induced across the disc (Fig.\,\ref{fig:b1}), which then evolves in accordance with the WKB relation (\ref{eq:Oms}). It winds up and propagates to the centre where it interferes with the bar, as follows from Fig.\,\ref{fig:SD}. While it remains relatively open, it produces a delayed response in the inner disc. This mechanism is similar to swing amplification observed in simulations of the stable Mestel disc that turns out to be surprisingly vital under the influence of minor dipole perturbations (T81). This interference, in some cases, leads to bar formation delay (B-1m, B-1-110), or advancement (B-1), depending on enhancement or attenuation of the central structure. Keeping the phase of the tidal wave
fixed, the interference pattern depends on the bar phase, which is illustrated by comparison of B-1 and B-1-110 runs. This basic mechanism is contaminated by additional random waves, which appear throughout the disc, also producing  delayed responses in the inner disc, regardless of any encounter.

\begin{figure*}
   \centering
   \includegraphics [width = 85mm]{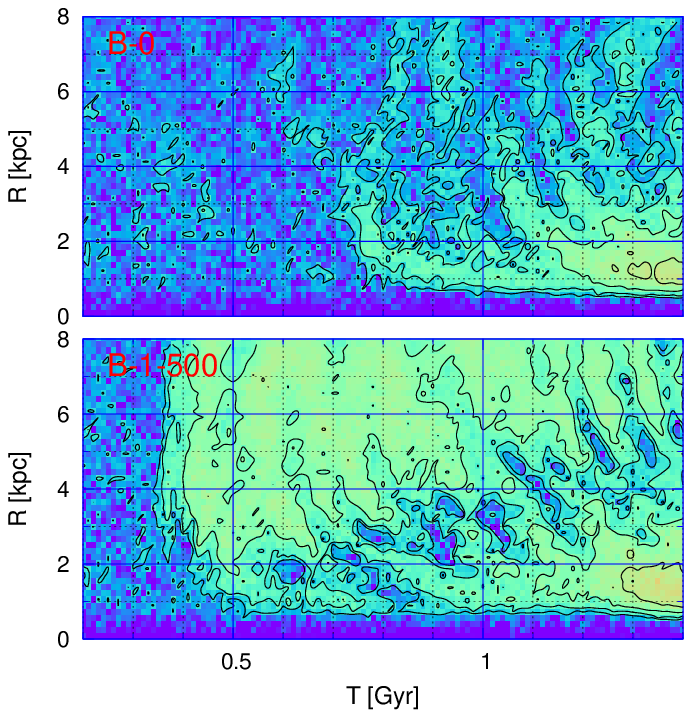} \hspace{1mm} \includegraphics [width = 85mm]{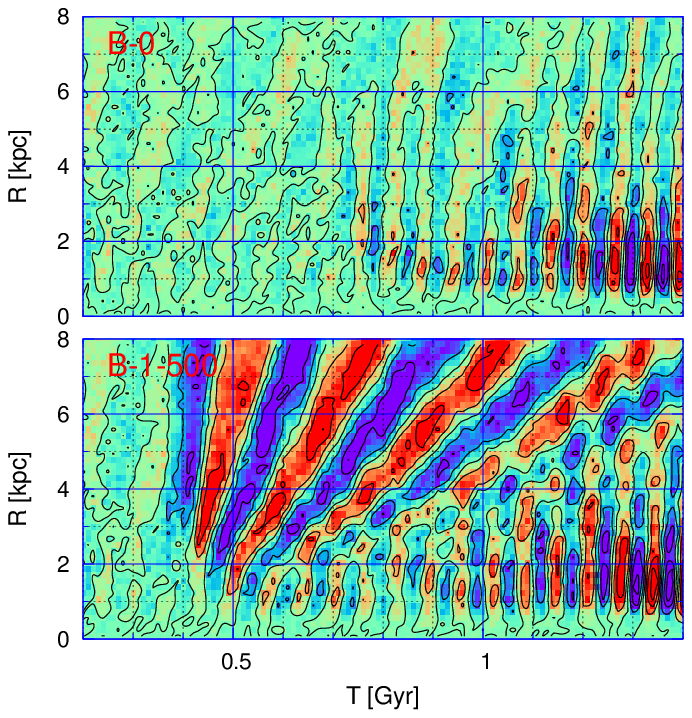}
   \caption{As in Fig.\,\ref{fig:SD} for B-0 and B-1-500 runs for the period $0.2<t<1.4$ Gyr.}
   \label{fig:SD500}
\end{figure*}

The effect is possible only if the encounter occurs after a seed bar is formed (after the amplitude jump at $t=t_1$) and before the exponential growth phase, $t\gtrsim t_2$. If the encounter occurs well before $t_1$, the tidal wave decays with no effect. This is proved by our B-1-500 run, in which the tidal wave occurs at $t \approx 400$ Myr, as follows from Fig.\,\ref{fig:SD500}. The bar formation time then coincides with one in the isolated run B-0. During the period of vulnerability, $t_1 \lesssim t \lesssim t_2$, the tidal wave amplitude is comparable to the amplitude of the seed bar, so the consequences of the interference are most evident. After $t_2$, the bar becomes strong so that it cannot be affected by relatively weak tidal waves.

It was known that the bar formation time depends on the instability properties of the galactic disc, which is determined by many parameters including the disc mass, velocity dispersion, dark matter halo, and a central matter concentration. This leads to uncertainties in predicting the bar age from simulations. The discussed phenomenon adds a scatter in the age of the bars. \citet{P16} showed that less unstable models have longer time between the first jump and the start of the exponential growth. This implies that at higher redshifts $z > 0.5$ satellites have more chances to interfere with the process of bar formation an thus the effect should be stronger. This analysis supports the hypothesis that all observed barred galaxies should be bar unstable in order to develop long-lasting bars.

\begin{acknowledgements}
We would like to thank the anonymous referee for their constructive comments and remarks which have greatly improved this paper. The main production runs were done on the {\tt MilkyWay} supercomputer, funded by the Deutsche Forschungsgemeinschaft (DFG) through the Collaborative Research Centre (SFB 881) ``The Milky Way System'' (subproject Z2), hosted and co-funded by the J\"ulich Supercomputing Center (JSC).

The special GPU accelerated supercomputer  {\tt laohu} funded by  NAOC/CAS and through the ``Qianren'' special foreign experts program of China (Silk Road Project), has been used for some of code development. We also used a smaller GPU cluster {\tt kepler} for data analysis, funded under the grants I/80 041-043 and I/81 396 of the Volkswagen Foundation. This work was supported by the Sonderforschungsbereich SFB 881 ``The Milky Way System'' (subproject A2 and A6) of the German Research Foundation (DFG). E.P. acknowledges financial support by the Russian Basic Research Foundation, grants 15-52-12387, 16-02-00649, and by the Basic Research Program OFN-15 `The active processes in galactic and extragalactic objects' of Department of Physical Sciences of RAS. P.B. acknowledges the special support by the NASU under the Main Astronomical Observatory GRID/GPU ``golowood'' computing cluster project.
\end{acknowledgements}

\end{document}